\documentclass[preprint,amsmath,amssymb,nofootinbib]{revtex4}
\usepackage[dvips]{graphics}
\usepackage{graphics}
\makeatletter
\renewcommand{\fnum@figure}{Fig.~ \thefigure}
\makeatother
\usepackage{pstcol}
\usepackage{amsfonts}
\usepackage{bm}
\usepackage{amsmath}
\usepackage{amssymb}
\usepackage{color}
\usepackage[all]{xy}
\usepackage{comment}
\usepackage{epstopdf,epsfig}
\def\be{\begin{equation}}
\def\ee{\end{equation}}
\def\bea{\begin{eqnarray}}
\def\eea{\end{eqnarray}}
\usepackage{float}
\usepackage{xcolor}
\usepackage{graphicx}
\usepackage{subcaption}
\usepackage{setspace}
\usepackage{microtype}
\usepackage[
    colorlinks=true,
    linkcolor=blue,     
    citecolor=red,      
]{hyperref}

\begin{document}

\title{Phase Transitions, Shadows, and Microstructure\\ of Kerr-anti-de Sitter Black Holes from Geometrothermodynamics\\}

\author{Jose Miguel Ladino$^1$, Carlos E. Romero-Figueroa$^1$, and Hernando Quevedo$^{1,2,3}$}
\email{\raggedright
miguel.ladino@correo.nucleares.unam.mx;
carlosed.romero@correo.nucleares.unam.mx;
quevedo@nucleares.unam.mx
}
\affiliation{\hspace{1cm}\\ \mbox{$^1$Instituto~de~Ciencias~Nucleares,~Universidad~Nacional~Aut\'onoma~de~M\'exico,}\\ \mbox{AP~70543,~Mexico~City,~Mexico}}
\affiliation{\mbox{$^2$Dipartimento~di~Fisica~and~Icra,~Universit\`a~di~Roma~“La~Sapienza”,~Roma,~Italy}}
\affiliation{\mbox{$^3$Al-Farabi~Kazakh~National~University,~Al-Farabi~av.~71,~050040~Almaty,~Kazakhstan}\\}

\date{\today}

\begin{abstract}
Using the formalism of geometrothermodynamics, we investigate the phase-transition structure and microstructure of the Kerr-anti-de Sitter black hole and show the relationship with its shadow structure. By treating the curvature radius as a thermodynamic variable, we ensure scaling consistency and model the system as quasi-homogeneous. In the canonical ensemble, we identify critical points and characterize both first- and second-order phase transitions independently of pressure. In the grand-canonical ensemble, we reveal a distinct phase structure, including the Hawking-Page transition. We derive analytical expressions for the Kerr-anti-de Sitter black hole shadow and its critical parameters, using shadow thermodynamics to construct asymmetric shadow profiles that capture the phase-transition structure. Finally, we show that the singularities of the geometrothermodynamic curvature  in the shadow align with divergences in thermodynamic response functions, confirming the correspondence between shadows, phase transitions, and microstructure.\\

\end{abstract}
\maketitle
\tableofcontents
\section{Introduction}
Recent observational achievements and projects, such as the first images of the supermassive black holes M87* and SgrA*, as well as future data on black hole shadows from the next-generation Event Horizon Telescope \cite{EventHorizon1,EventHorizon2,EventHorizon3} and the Black Hole Explorer \cite{Johnson}, have significantly increased interest in exploring black hole properties within general relativity and its extensions. In this work, we focus on one of the most fundamental generalizations of a spherically symmetric black hole in Einstein's theory that incorporates angular momentum and a negative cosmological constant: the well-known Kerr-Anti-de Sitter (Kerr-AdS) black hole, which is the most general black hole solution of Einstein gravity theory with cosmological constant. 

Meanwhile, differential geometry has proven to be a powerful tool for examining thermodynamic systems from a novel perspective. In particular, the framework of thermodynamic geometry employs Hessian metrics to describe the equilibrium space of thermodynamic systems as Riemannian manifolds \cite{amari2012differential, weinhold1976metric, ruppeiner1979thermodynamics}. Notably, the Ruppeiner metric has been extensively used in black hole thermodynamics \cite{ruppeiner1979thermodynamics}. More recently, the formalism of geometrothermodynamics (GTD) \cite{quevedo2007geometrothermodynamics} was introduced to ensure the Legendre invariance of classical thermodynamics within the geometric formulation of the equilibrium space. A vast number of recent studies have employed thermodynamic geometry to examine phase transitions, stability, and the microstructure of various rotating black holes \cite{wei2021general, ref39, Banerjee, ref20, Zangeneh_2018, Dehyadegari, Sahay, Hazarika, quevedo2008geometrothermodynamics, Gogoi, larranaga2012geometric, ref54, wei2020extended, ruppeiner2008thermodynamic}. In particular, Ruppeiner's geometric approach has previously been used to study the phase structure and microstructure of the Kerr-AdS black hole, revealing a configuration of first- and second-order phase transitions \cite{ Banerjee, ref20, wei2021general, Zangeneh_2018, Dehyadegari, Sahay, Hazarika}. However, some difficulties arise in choosing the thermodynamic variable associated with the cosmological constant, which is why a form of pressure is introduced in the extended black hole thermodynamics approach \cite{wei2021general, Zangeneh_2018, Dehyadegari, Sahay, gunasekaran2012extended, altamirano2014kerr}. In this work, we employ the GTD approach to investigate the thermodynamic properties of the Kerr-AdS black hole, treating the curvature radius \( l \) as a thermodynamic variable without introducing any notion of pressure.     

In \cite{Wei_2019}, one of the first analyses of the relationship between the unstable circular photon orbit and the phase transition of the rotating Kerr-AdS black hole was conducted. Numerous studies within alternative gravity theories have explored the thermodynamic implications of photon orbits and black hole shadows \cite{ZhangGuo, Luo, GuoLi, Belhaj, WangRuppeiner, Li, Kumar, Du, Zheng, He, ref20, ref21}. This line of research has led to the development of the shadow thermodynamics approach, which examines black hole thermodynamics based on the features of their shadows. A key motivation for this methodology is the realization that certain measurable and observable features of black hole shadows reliably indicate thermodynamic behavior, including phase transitions. Studies in \cite{WangRuppeiner, ref20, ref21} reveal that, within the Ruppeiner framework, shadow features provide insights into the phase structure and microstructure of AdS black holes. \\

In a recent work \cite{Ladino:2024ned}, we studied the thermodynamic properties of the Reissner-Nordström-AdS (RN-AdS) black hole using the shadow thermodynamics approach and the GTD formalism. We derived explicit expressions for the shadow radius in terms of the event horizon, photon sphere, and observer radii, analyzing the phase structure and microscopic properties. In this work, we once again use the condition that the Kerr-AdS black hole behaves as a quasi-homogeneous thermodynamic system to ensure the consistency of the GTD approach. As a result, the cosmological constant, expressed in terms of the curvature radius $l$, becomes a thermodynamic variable.  This paper is organized as follows. In Sec. \ref{sec:shadow}, we review the main aspects of Kerr-AdS black hole and its shadows, emphasizing the influence of rotation and the observer’s location. In Sec. \ref{sec:thermo}, we analyze the thermodynamic properties and phase structure of Kerr-AdS black holes, considering both the canonical and grand-canonical ensembles.  In Sec. \ref{GTDsection}, we apply the formalism of GTD to study in detail the phase transitions and microstructure of the Kerr-AdS spacetime, treating the curvature radius as a thermodynamic variable. Furthermore, in Sec. \ref{sec:shtd}, we investigate the thermodynamic black hole properties trough its shadows, exploring their relation to phase transitions and extracting thermodynamic information from shadow size and asymmetric profiles, extending the GTD framework to the context of rotating black hole shadows. Finally, in Sec. \ref{sec:conclusions}, we summarize our main results.

\section{Shadows of Kerr-AdS Black Holes}
\label{sec:shadow}
The gravitational Einstein-Hilbert action in AdS spacetime is given by
\begin{equation}
 \label{accion1}
S_g= \frac{1}{16 \pi G}\int d^{4}x\sqrt{-g}\left(R-2\Lambda\right ).
\end{equation}
The general solution representing a rotating Kerr-AdS black hole was found by Carter \cite{carter1968hamilton} and in Boyer-Lindquist coordinates can be expressed as \cite{dias2012kerr}
\begin{equation}
    ds^2=-\frac{\Delta_r}{\Sigma^2}\Big(dt-\frac{a \sin^2 \theta}{\Xi}d\phi\Big)^2+\frac{\Sigma^2}{\Delta_r}dr^2+\frac{\Sigma^2}{\Delta_\theta}d\theta^2+\frac{\Delta_\theta}{\Sigma^2}\sin^2 \theta\Big(a dt-\frac{r^2+a^2}{\Xi}d\phi^2\Big)^2,\label{metric1}
\end{equation}
where
\begin{equation}
\Delta_r = (r^2 + a^2) \left( 1 + \frac{r^2}{l^2} \right) - 2mr, \quad
\Xi = 1 - \frac{a^2}{l^2}, \quad
\Delta_\theta = 1 - \frac{a^2}{l^2} \cos^2 \theta, \quad
\Sigma^2 = r^2 + a^2 \cos^2 \theta.\label{metric22}
\end{equation}
This solution satisfies $R_{\mu \nu} = -3g_{\mu \nu}/l^2$ and asymptotically approaches the AdS space with a curvature radius $l$. The cosmological constant is given by $
\Lambda = -3/l^2$. Furthermore, the ADM mass $M$ and the angular momentum
$J$ of the Kerr-AdS black hole are related to the parameters $m$ and $a$ as \cite{caldarelli2000thermodynamics}
\begin{equation}
\begin{aligned}
M=\frac{ m}{ \Xi^2} \quad \text { and } \quad J=\frac{ a m}{ \Xi^2}. 
\end{aligned}
\label{MandJ}
\end{equation} 
The event horizon is located at $r = r_+$, the largest real root of $\Delta_r=0$. The rotational parameter $a$ is constrained by $a < l$. Solutions where this bound is saturated no longer describe black holes. As $a$ approaches $l$ with fixed $r_+$, both the mass and angular momentum of the black hole become unbounded, and the circumference at the equator grows without limit. The  Hawking temperature and entropy for a Kerr-AdS black hole are given by

\begin{equation}
T = \frac{\kappa}{2\pi} = \frac{r_+ \left(1 + \frac{a^2}{l^2} + \frac{3r_+^2}{l^2} - \frac{a^2}{r_+^2}\right)}{4\pi(r_+^2 + a^2)},\quad S = \frac{A}{4} = \frac{\pi(r_+^2 + a^2)}{1 - \frac{a^2}{l^2}}, \label{temp}
\end{equation}
respectively, where $\kappa$ is the surface gravity at the event horizon and $A$ is the horizon area\footnote{In natural units where $\hbar=c=\kappa_B=G=1$.} of the black hole. The Kerr-AdS black hole exhibits a regular extremal state where the temperature reaches zero, but the entropy remains finite. The extremality conditions $T = 0$ and $\Delta_r(r_+) = 0$ allow the parameters $a = a_{\text{ext}}$ and $m = m_{\text{ext}}$ to be expressed as functions of $l$ and $r_+$:
\begin{equation}
a_{\text{ext}} = r_+ \sqrt{\frac{3r_+^2 + l^2}{l^2 - r_+^2}},\quad m_{\text{ext}} = \frac{r_+ \left( 1 + \frac{r_+^2}{l^2} \right)^2}{1 - \frac{r_+^2}{l^2}}. \label{extremalcondition}
\end{equation}
Note that only black holes with $r_+ <l/ \sqrt{3}$ can reach zero temperature due to the constraint $a <l$. On the other hand, in the spacetime of Eq.~ \eqref{metric1}, null geodesic motion can be described using the Hamilton-Jacobi formalism \cite{chandrasekhar},
\begin{equation}
\begin{aligned}
\frac{\partial \mathcal{S}}{\partial \tau}+\frac{1}{2} g^{\mu \nu} \frac{\partial \mathcal{S}}{\partial x^\mu} \frac{\partial \mathcal{S}}{\partial x^\nu}=0  \quad \text { and } \quad \mathcal{S}=-E t+L \phi+\mathcal{S}_r(r)+\mathcal{S}_\theta(\theta).
\end{aligned}
\label{HamiltonJacobi}
\end{equation} 

Here, $\tau$ is an affine parameter, $x^\mu$ denotes $(t, r, \theta, \phi)$, and $\mathcal{S}$ represents Hamilton’s principal function, which can be shown to be separable. The constants of motion, energy $E$ and angular momentum $L$, are associated with the Killing vectors $\partial / \partial t$ and $\partial / \partial \phi$, respectively. Thus, solving Eq.~ \eqref{HamiltonJacobi}, the resulting equations that describe the light trajectories are 
\begin{align}
 \Sigma \frac{d t }{d \tau}&=\frac{\Xi^2 }{\Delta_r}\left[\left(r^2+a^2\right) E-a L\right]\left(r^2+a^2\right)-\frac{a \Xi^2}{\Delta_\theta}\left(a E \sin ^2 \theta-L\right), \label{teq}\\
R(r) \equiv \Sigma^2 \left(\frac{d r }{d \tau}\right)^2&=\Xi^2\left[\left(r^2+a^2\right) E-a L\right]^2-\mathcal{Q} \Delta_r, \label{req}\\
\Theta(\theta) \equiv \Sigma^2 \left(\frac{d \theta }{d \tau}\right)^2&=-\frac{\Xi^2}{\sin ^2 \theta}\left(a E \sin ^2 \theta-L\right)^2+\mathcal{Q} \Delta_\theta,\label{Thetaeq} \\
\Sigma \frac{d \phi }{d \tau}&=\frac{a \Xi^2 }{\Delta_r}\left[\left(r^2+a^2\right) E-a L\right]-\frac{\Xi^2 }{\Delta_\theta \sin ^2 \theta}\left(a E \sin ^2 \theta-L\right), \label{phieq}
\end{align}
where $\mathcal{Q}$ is the separable Carter constant. In the photon region, light rays are forced by gravity to travel in orbits. Unlike the static case, for Kerr-AdS black holes, unstable circular photon orbits exist only in the equatorial plane and come in two types: the prograde (corotating) orbit lies at a smaller radius $r_s^-$, while the retrograde (counterrotating) orbit is at a larger radius $r_s^+$ \cite{Perlick_2022}. The shape of a black hole shadow is determined by the constant values of $r=r_s$, the radius of circular photon orbits, which must satisfy the following conditions
\begin{equation}
\begin{aligned}
\left.R(r)\right|_{r=r_s}=\left. \partial_r R(r)\right|_{r=r_s}=0,
\end{aligned}
\label{conditions1}
\end{equation}
and
\begin{equation}
\begin{aligned}
\Theta(\theta) \geq 0 \quad \text { for } \quad  \theta \in[0, \pi] .
\end{aligned}
\label{conditions2}
\end{equation}
To simplify the analysis, we introduce two impact parameters $\xi \equiv L / E$ and $\eta \equiv \mathcal{Q} / E^2$, and we set $R / E^2 \rightarrow R$ and $\Theta / E^2 \rightarrow \Theta$, from now on. Then, solving Eqs.\eqref{conditions1} for the impact parameter, we obtain
\begin{equation}
\begin{aligned}
 \xi=\frac{r_s^2+a^2}{a}- \frac{4 r_s\Delta_r\left(r_s\right)}{a\Delta_r^{\prime}\left(r_s\right)}\quad \text { and } \quad \eta=\frac{16 r_s^2 \Xi^2 \Delta_r\left(r_s\right)}{\left[\Delta_r^{\prime}\left(r_s\right)\right]^2}.
\end{aligned}
\label{impactfactors}
\end{equation}
These impact parameters are functions of \( r_s \), which in turn depends on the black hole parameters \( \{m, l, a\} \). Therefore, \(\xi\) and \(\eta\) ultimately depend only on \( \{m, l, a\} \). From Eqs.~ \eqref{Thetaeq}, \eqref{conditions2}, and \eqref{impactfactors}, we find the  polynomial
\begin{equation}
\begin{aligned}
 4\left(a^3-a l^2\right)^2\left[r_s^4+l^2 r_s(r_s-2 m)+a^2\left(1^2+r_s^2\right) \right] &\\
-l\left[l^3 r_s(r_s-3 m)+a^2 l\left(2 l^2+r_s^2\right) \right] &^2=0.
\end{aligned}
\label{photon1}
\end{equation}
The two positive real solutions of this polynomial define the boundaries of the photon region in the equatorial plane, $r_s^- \leq r_s \leq r_s^+$, which determines the shape of the critical curve of the black hole shadow.\\

When the cosmological constant is considered, it has been established that the observer's distance within the domain of outer communication must be specified. So, special caution must be taken with spacetimes that are not asymptotically flat, as in the case of the Kerr-AdS black hole. To analyze the shadow observed from a finite distance, we employ the method introduced in \cite{Griffiths_Podolský_2009, Grenzebach1, Grenzebach2} and later applied in \cite{Belhaj_2022-2, Belhaj_2022, Fathi_2021, grenzebach2016shadow, Zubair_2022, Belhaj2021-2, Belhaj_2021, Perlick_2017, Chowdhuri_2021, Perlick_2022, Haroon:2022ehm, Haroon_2019}. We place the observer in an orthonormal tetrad frame \( \left(e_0, e_1, e_2, e_3\right) \) at \( (r = r_0, \theta = \theta_0) \), such that \cite{Haroon_2019}
\begin{equation}
\begin{aligned}
e_0 &=\left.\frac{\Xi^2}{\sqrt{\Delta_r \Sigma}}\left[\left(r^2+a^2\right) \partial_t+a \partial_\phi\right]\right|_{r=r_0, \theta=\theta_0}, \quad\hspace{0.55cm} e_1=\left.\sqrt{\frac{\Delta_\theta}{\Sigma}} \partial_\theta\right|_{r=r_0, \theta=\theta_0}, \\
e_2 &=-\left.\frac{\Xi^2}{\sqrt{\Delta_\theta \Sigma} \sin \theta}\left(\partial_\phi+a \sin ^2 \theta \partial_t\right)\right|_{r=r_0, \theta=\theta_0}, \quad 
e_3=-\left.\sqrt{\frac{\Delta_r}{\Sigma}} \partial_r\right|_{r=r_0, \theta=\theta_0}.
\end{aligned}
\label{tetrads}
\end{equation}
The vector $e_3$ points in the spatial direction toward the center of the black hole. The timelike vector $e_0$ represents the four-velocity of the observer, while $e_0 \pm e_3$ are tangential to the principal null congruences. Then, we define 
$\lambda(s) = 
(r(s), \theta(s), \phi(s), t(s))$ as the coordinates of an arbitrary light ray. At the observer's position, the tangent vector can be represented in two equivalent forms (more details in \cite{Grenzebach1, Grenzebach2, Haroon_2019}): one in terms of the coordinate basis and the other using the previously introduced tetrad along with the celestial coordinates $\rho$ and $\sigma$ as
\begin{align}
\dot{\lambda}=&\dot{r} \partial_r+\dot{\theta} \partial_\theta+\dot{\phi} \partial_\phi+\dot{t} \partial_t , \label{tangent1}\\
\dot{\lambda}=&\mu\left(-e_0+\sin \rho \cos \sigma e_1+\sin \rho \sin \sigma e_2+\cos \rho e_3\right),
\label{tangent2}
\end{align}
where $\mu$ is a scalar factor that can be calculated from
the two previous expressions as
\begin{equation}
\mu=\left.\Xi^2 \frac{a L-E\left(r^2+a^2\right)}{\sqrt{\Delta_r \Sigma}}\right|_{r=r_0, \theta=\theta_0}.
\end{equation}
Using the expressions for \(\dot{r}\) and \(\dot{\phi}\) from Eqs.~ \eqref{req} and \eqref{phieq}, the tetrad components from Eqs.~ \eqref{tetrads}, and the impact parameters \(\xi\) and \(\eta\) from Eqs.~ \eqref{impactfactors}, we can determine the celestial coordinates \(\rho\) and \(\sigma\) by matching the coefficients of \(\partial_r\) and \(\partial_\phi\) in Eq.~ \eqref{tangent1} with those in Eq.~ \eqref{tangent2}. We obtain 
\begin{equation}
\begin{aligned}
\sin \rho=\pm\sqrt{1-\frac{1}{\Xi^2}+\frac{16 r_s^2 \Delta_r(r_0)  \Delta_r\left(r_s\right)}{\big[4 r_s \Xi \Delta_r\left(r_s\right)+\left(r_0^2-r_s^2\right) \Xi \Delta_r^{\prime}\left(r_s\right)\big]^2}}
\end{aligned}
\label{celestial1}
\end{equation}
and 
\begin{equation}
\begin{aligned}
\sin \sigma=\frac{\Xi \csc \theta_0  \sqrt{\Delta_r(r_0)}\Big\{\big[a^2+2 r_s^2+a^2 \cos \left(2 \theta_0\right)\big] \Delta_r^{\prime}\left(r_s\right)-8 r_s \Delta_r\left(r_s\right)\Big\}}{2 a \sqrt{\Delta_\theta\left( \theta_0\right)} \sqrt{16 r_s^2 \Delta_r(r_0)  \Delta_r\left(r_s\right)+\left(\Xi^2-1\right)\left(4 r_s \Delta_r\left(r_s\right)+\left(r_0^2-r_s^2\right) \Delta_r^{\prime}\left(r_s\right)\right)^2}}
\end{aligned}
\label{celestial2}
\end{equation}

These celestial coordinates, \(\rho\) and \(\sigma\), are also functions of \( r_s \). Therefore, they depend solely on the black hole parameters and the observer's position, namely \( \{m, l, a, r_0, \theta_0\} \). Thus, the critical curve of the black hole shadow can be represented onto a plane using a stereographic projection from the celestial sphere in cartesian coordinates as
\begin{equation}
\begin{aligned}
& x=-2 \tan \left(\frac{\rho}{2}\right) \sin \sigma,\quad \text { and } \quad   y=-2 \tan \left(\frac{\rho}{2}\right) \cos \sigma.
\label{cartecian}
\end{aligned}
\end{equation}
\begin{figure}[H]
  \centering
  \begin{subfigure}{0.48\textwidth}
  \caption*{(a) \hspace{0.5cm}  $m=1$, $l=10$, $r_0=100$, $\theta_0=\pi/2$}
    \includegraphics[width=\linewidth]{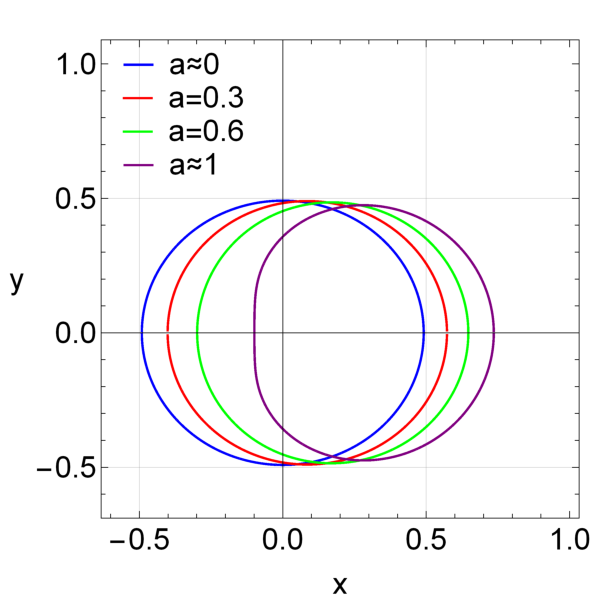}
  \end{subfigure}
  \hfill
  \begin{subfigure}{0.48\textwidth}
  \caption*{(b) \hspace{0.5cm}  $m=1$, $a=0.6$, $r_0=100$, $\theta_0=\pi/2$}
    \includegraphics[width=\linewidth]{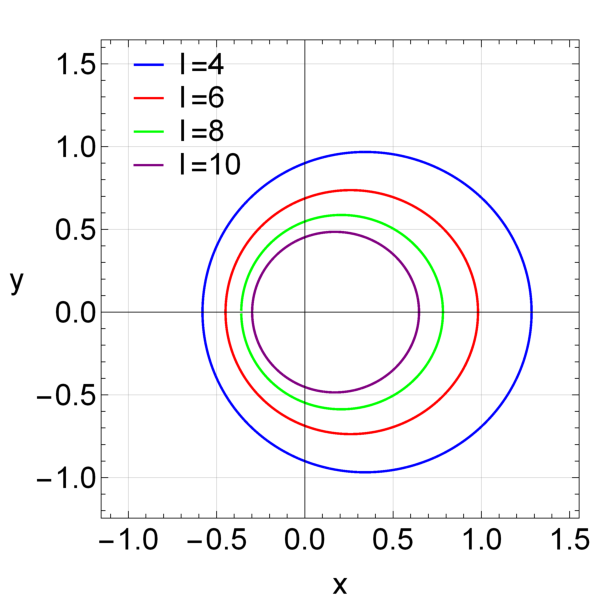}
  \end{subfigure}
  \vspace{0.5cm}
    \centering
  \begin{subfigure}{0.48\textwidth}
  \caption*{(c) \hspace{0.5cm}  $m=1$, $a =0.8$, $l=10$, $\theta_0=\pi/2$}
    \includegraphics[width=\linewidth]{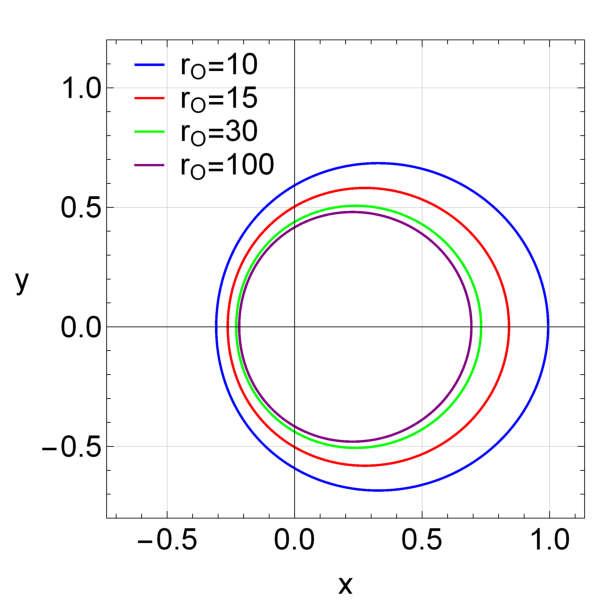}
  \end{subfigure}
  \hfill
  \begin{subfigure}{0.48\textwidth}
  \caption*{(d) \hspace{0.5cm}  $m=1$, $a =0.8$, $l=10$, $r_0=100$}
    \includegraphics[width=\linewidth]{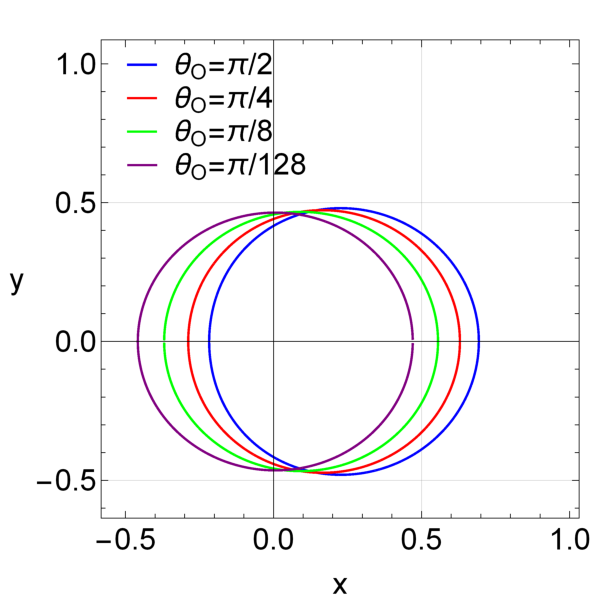}
  \end{subfigure}
  \caption{Shadow behavior of the Kerr-AdS black hole for different values of \( a \), \( l \), \( \theta_0 \), and \( r_0 \). }
  \label{fig:figura0_completa}
\end{figure}
To examine how the parameters affect the black hole shadow, we can associate them with observables that are measurable and valuable for testing black hole characteristics. We will use two key observables parameters: the radius of the reference circle for the distorted shadow, $R_{sh}$, and the deviation of the shadow’s left edge from the boundary of the reference circle, $\delta_{sh}$ \cite{Hioki_2009}. We introduce the top, bottom, right, and left points of the shadow as $(x_t,y_t)$, $(x_b,y_b)$, $(x_r,0)$ and $(x_l,0)$, respectively, and denote the point $(x_l^{\prime},0)$ as the leftmost edge of the reference circle. With these definitions, the observables can be written as 
\begin{equation}
R_{sh} \equiv \frac{\left(x_t-x_r\right)^2+y_t^2}{2\left|x_r-x_t\right|}, \quad \text { and } \quad \delta_{sh} \equiv \frac{\left|x_l-x_l^{\prime}\right|}{R_{sh}} .
\label{radiusshadow}
\end{equation}
\begin{figure}[H]
  \centering
  \begin{subfigure}{0.46\textwidth}
  \caption*{(a) $\hspace{6.2cm}$ $\vspace{0.4cm}$ }
    \includegraphics[width=\linewidth]{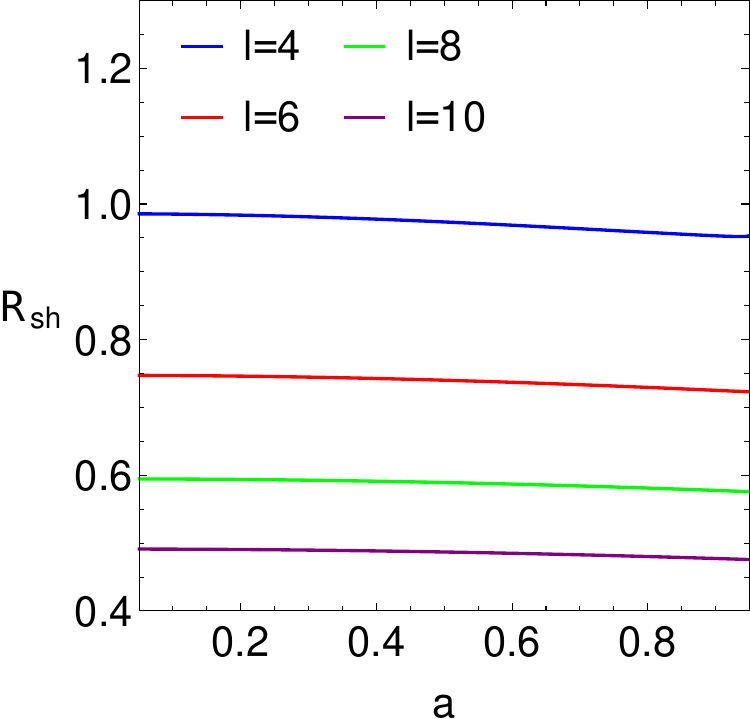}
  \end{subfigure}
  \hfill
  \begin{subfigure}{0.46\textwidth}
  \caption*{(b) $\hspace{6.2cm}$ $\vspace{0.2cm}$}
    \includegraphics[width=\linewidth]{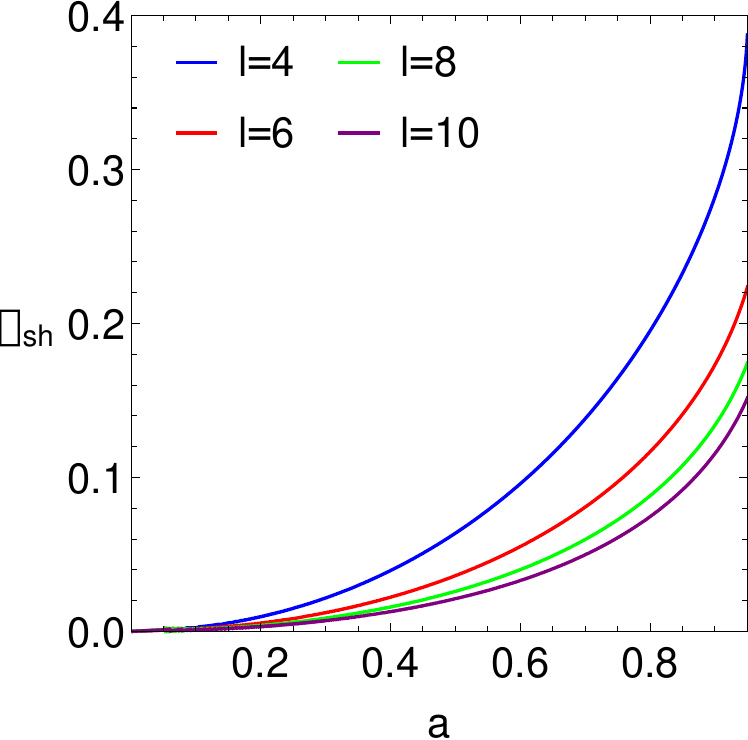}
  \end{subfigure}
  \caption{ (a) Shadow radius $R_{sh}$ and (b) deviation  $\delta_{sh}$ as functions of the spin parameter $a$, for various values of $l$ (with $m=1$, $r_0=100$ and $\theta_0=\pi/2$).}
  \label{fig:shadow1}
\end{figure}
In Fig.~\ref{fig:figura0_completa}, we present the shadow of a Kerr-AdS black hole as a function of the parameters $a$, $l$, $r_0$, and $\theta_0$. As expected, the critical curve loses circularity when the black hole's rotation is present, and as the curvature radius $l$ increases, the shadow shrinks, approaching that of a Kerr black hole in asymptotically flat spacetime. The shape of a rotating black hole shadow is mainly determined by its spin parameter \( a \). As \( a \) increases, the shadow becomes more asymmetric due to the "dragging" of light rays, a well-known effect in the Kerr metric. In the extremal case, when \( a \approx a_{\text{ext}} \), the shadow takes on a distinct D-shape, as first described by Bardeen in 1973 \cite{Bardeen}. The shadow remains nearly circular when the observer is in an almost polar position, as seen when \( \theta_0 \) approaches $0$. However, for highly charged black holes or in certain generalized spacetimes, the shadow remains nearly circular even at high spin values, including in the extremal case \cite{grenzebach2016shadow}. On the other hand, Fig.~\ref{fig:shadow1} illustrates the shadow radius $R_{sh}$ and deviation $\delta_{sh}$ as functions of the spin parameter $a$ for different values of $l$. In general, as the curvature radius $l$ increases, both $R_{sh}$ and $\delta_{sh}$ decrease. Further, as $a$ grows, $R_{sh}$ decreases while $\delta_{sh}$ increases, results obtained previously in  \cite{Hioki_2009, Papnoi_2014, Jusufi_20192, Pantig_2023}.


\section{Thermodynamics and Phase Transition Structure}\label{sec:thermo}

To perform a thermodynamic analysis based on scaling arguments, implying that the corresponding thermodynamic system is quasi-homogeneous,  it is essential to treat the spacetime curvature radius \( l \) as an independent thermodynamic variable \cite{romero2024extended1,Ladino:2024ned}. This approach enables the consideration of ensembles where \( l \) is allowed to fluctuate. To the best of our knowledge, this is the first study to explore the thermodynamic effects by explicitly including $l$ as a thermodynamic variable in the Kerr-AdS system. Other related works in the enthalpy representation introduce the notion of pressure (see \cite{wei2021general,gunasekaran2012extended,altamirano2014kerr, Zangeneh_2018, Dehyadegari, Sahay,karch2015holographic,mancilla2024generalized}). Consequently, the parameters \( (S, l, J) \) can be regarded as a complete set of independent "extensive" variables defining the black hole's fundamental relation, which encodes the full thermodynamic state of the Kerr-AdS black hole, as given in \cite{caldarelli2000thermodynamics}

\begin{equation}
M(S,l,J) = \sqrt{J^2 \left(\frac{\pi}{S}  + \frac{1}{l^2} \right) + \frac{S^3}{4\pi^3} \left( \frac{\pi}{S} + \frac{1}{l^2} \right)^2}.\label{fundame1}
\end{equation}
The above relation holds for classical black holes and in the limit $l \rightarrow \infty$ this formula reduces to the usual Smarr formula for asymptotically flat Kerr solutions \cite{smarr1973mass}. Using the fundamental equation \eqref{fundame1}, we can express the "intensive" thermodynamic quantities as equations of state, i.e.,

\begin{align}
T (S,l,J) &\equiv \frac{\partial M}{\partial S} \bigg|_{lJ} = \frac{1}{8\pi M} \left( 1 -\frac{4\pi^2J^2}{S^2}+ \frac{4S}{\pi l^2} + \frac{3S^2}{\pi^2 l^4} \right), \label{tempkerr}\\
\Omega (S,l,J)&\equiv\frac{\partial M}{\partial J} \bigg|_{S l} = \frac{\pi J}{M S} \left( 1 + \frac{S}{\pi l^2} \right),\\
\Psi (S,l,J) &\equiv\frac{\partial M}{\partial l} \bigg|_{S J}= -\frac{1}{M l^3}\left[ J^2 + \frac{S^2 \left(l^2 \pi + S\right)}{2 l^2 \pi^3}
 \right],
\end{align}
where the effective angular velocity \( \Omega \) that appears as  thermodynamic parameter is given by the difference between the angular velocities at the event horizon \( \Omega_H \) and at infinity\footnote{ Note that, in contrast to the asymptotically flat Kerr black hole, \( \Omega_\infty \)  $\neq 0$.
} \( \Omega_\infty \) \cite{caldarelli2000thermodynamics}.
Employing these thermodynamic quantities, one can verify that the following first law holds
\be
dM = TdS + \Omega dJ + \Psi dl. 
\ee

\subsection{Canonical ensemble}

In the canonical ensemble, the natural control parameters are the angular momentum $J$ and the curvature radius $l$. Consequently, the black hole temperature, as given in Eq.~ \eqref{tempkerr}, can be interpreted as an equation of state for studying the phase structure of the Kerr-AdS solution. Moreover, when the parameters $(J, l)$ are held fixed, the phase transition of the system is effectively characterized by the heat capacity $C_{lJ}$, which is defined as

\begin{align}
 C_{lJ}&\equiv T\Big(\frac{\partial S}{\partial T}\Big)_{lJ} \label{temp4}\\
 &=\frac{2 S\left(l^2 \pi+S\right)\big[4 J^2 l^2 \pi^3+S^2\left(l^2 \pi+S\right)\big] \big[S^2\left(l^2 \pi+S\right)\left(l^2 \pi+3 S\right)-4 J^2 l^4 \pi^4\big]}{\left(l^2 \pi-3 S\right) S^4\left(l^2 \pi+S\right)^3+24 J^2 l^2 \pi^3 S^2\left(l^2 \pi+S\right)^2\left(l^2 \pi+2 S\right)+16 J^4 l^6 \pi^7\left(3 l^2 \pi+4 S\right)}. \notag
\end{align}
\begin{figure}[H]
  \centering
  \begin{subfigure}{0.45\textwidth}
  \caption*{(a) $\hspace{6.2cm}$ $\vspace{0.2cm}$ }
    \includegraphics[width=\linewidth]{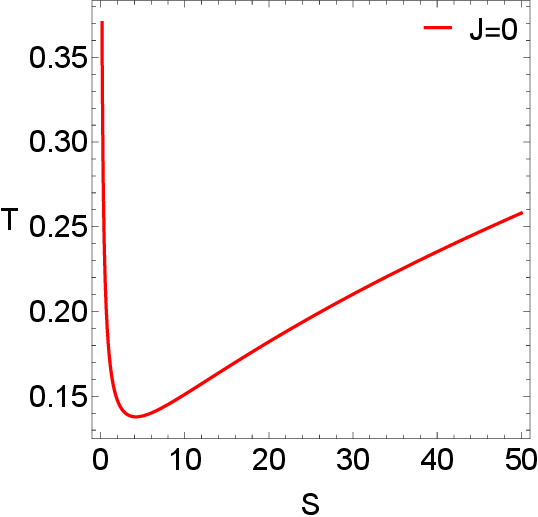}
  \end{subfigure}
  \hfill
  \begin{subfigure}{0.47\textwidth}
  \caption*{(b) $\hspace{6.2cm}$ $\vspace{0.2cm}$}
    \includegraphics[width=\linewidth]{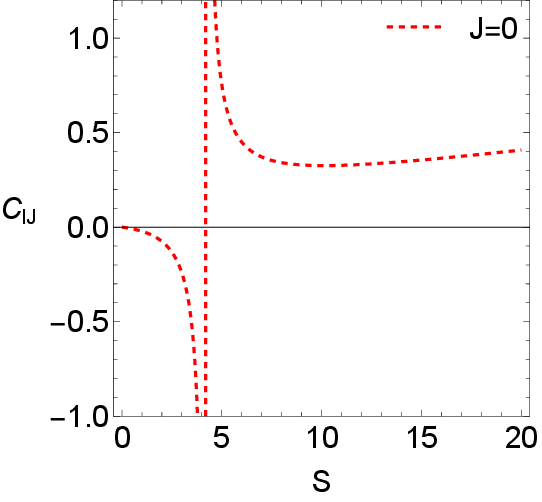}
  \end{subfigure}
  \caption{ Temperature $T$ and normalized heat capacity $C_{lJ}$ of the Kerr-AdS black hole in the Schwarzschild-AdS limit (\( J = 0 \)) for fixed \( l = 2 \), as a function of entropy. (a) $T$ shows a local minimum marking the transition from an unstable to a stable configuration. (b) $C_{lJ}$ diverges at this point, indicating a change in stability conditions.\label{temp_j_zero}}
\end{figure}
The temperature, which must remain a positive quantity, reaches its minimum value at $S_0=S_0(l,J)=S(T=0)$, this is
\begin{align}
& S_0=  \frac{\pi l}{6}\left(\sqrt{\frac{\left(l^2+J_{0}\right)^2-144 J^2}{J_{0}}}-2 l+\sqrt{4 l^2+\frac{144 J^2-l^4}{J_{0}}-J_{0}-\frac{4 l^3 \sqrt{J_{0}}}{\sqrt{\left(l^2+J_{0}\right)^2-144 J^2}}}\right),\label{smintemperat}\\
& J_{0} \equiv \left(l^6-432 J^2 l^2+12 J \sqrt{20736 J^4+864 J^2 l^4-3 l^8}\right)^{1 / 3}\notag. 
\end{align}
The heat capacity exhibits a divergence precisely at the points where the temperature reaches an extremum, denoted by \( T = T_{\text{extr}} \), which occurs when the derivative of the temperature with respect to the entropy vanishes, i.e., \( \partial T/\partial S = 0 \). According to the standard Ehrenfest scheme \cite{callen1998thermodynamics}, and within the context of black hole thermodynamics \cite{davies1978thermodynamics}, these points correspond to a phase transition in the thermodynamic system. These critical curves, are given by 
 $S_{extr}=S_{extr}(l,J)=S(T=T_{extr})$, where  $T_{min,max}= T(S_{extr})$. For black holes to undergo a second-order phase transition, there must exist a location where the local extrema merge into a single inflection point
\cite{kubizvnak2017black}. Using Eq.~ \eqref{tempkerr} and the conditions $\partial_S T =\partial_{SS} T=0$, the critical points of the system are obtained. To proceed analytically, only contributions up to order $ \mathcal{O}(J^2)$ will be considered. Under this approximation, the following critical values can be obtained (for the explicit forms of the approximate $S_{extr}$ and critical values, see Eqs.~ \eqref{sextrem} and \eqref{critical1apen} in Appendix \ref{apenA})
\begin{equation}
\begin{aligned}
 l_c &  \approx 6.4407 \sqrt{J}, \quad  \quad S_c \approx28.3114 J,\quad \quad   M_c \approx 1.8637 \sqrt{J}, \\
 T_c &\approx\frac{0.0419}{\sqrt{J}},\quad \quad \quad   r_{hc}\approx2.9430 \sqrt{J}.
\end{aligned}
\label{critical1}
\end{equation}
Numerically, we can also find that $l_c \approx 6.4640 \sqrt{J}$,  $ S_c \approx28.7189 J$,  $ M_c \approx 1.8783 \sqrt{J}$,  
 $T_c \approx 0.0417 J^{-1/2}$, and $  r_{hc}\approx2.9658 \sqrt{J}$. These critical quantities are consistent with the results previously obtained in \cite{Wei_2019, wei2016analytical,gunasekaran2012extended,wei2021general, Xiong}.
\begin{figure}[H]
  \centering
  
  \begin{subfigure}{0.33\textwidth}
  \caption*{ $ J=0.3$}
    \includegraphics[width=\linewidth]{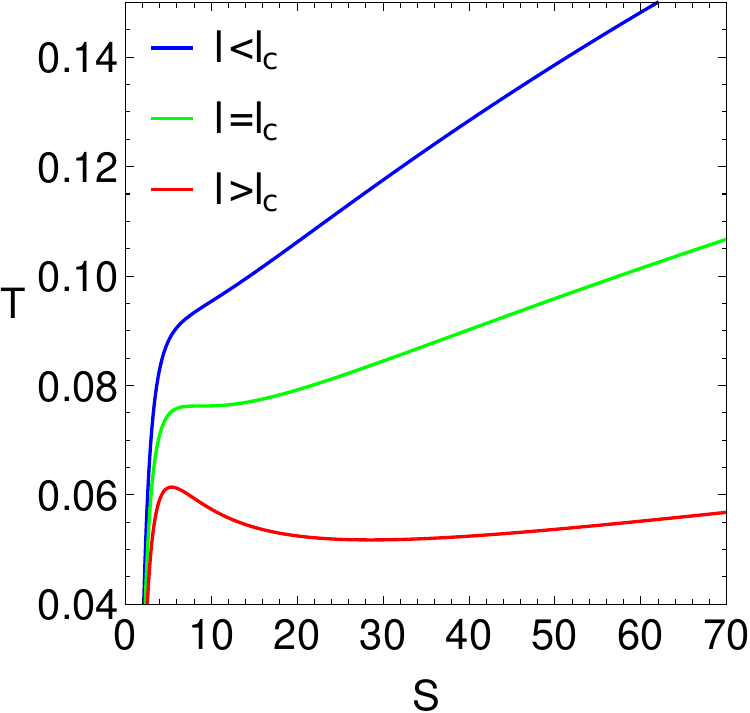}
  \end{subfigure}
  \hfill
  \begin{subfigure}{0.32\textwidth}
      \caption*{ $ J=0.6$}
    \includegraphics[width=\linewidth]{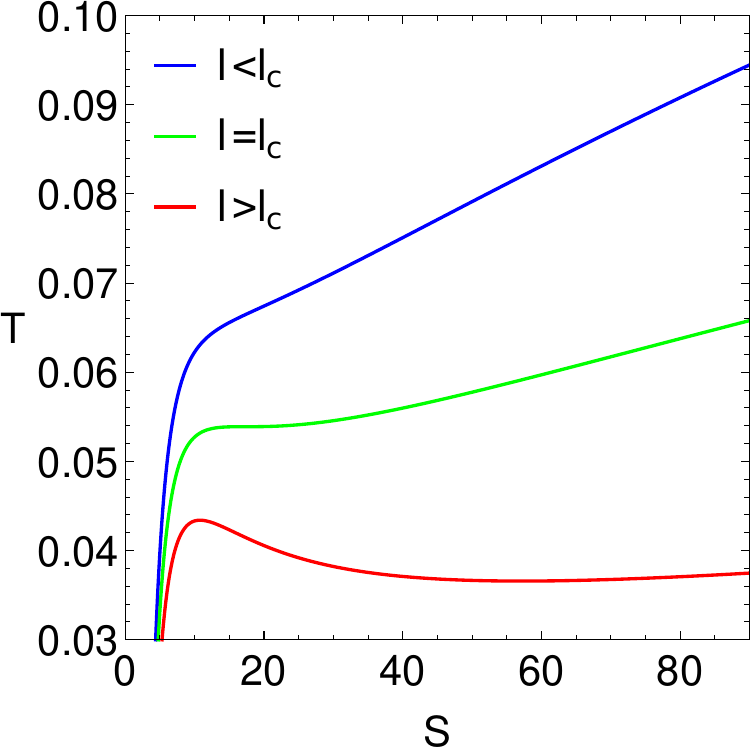}
  \end{subfigure}
  \hfill
  \begin{subfigure}{0.33\textwidth}
      \caption*{ $ J= 1$}
    \includegraphics[width=\linewidth]{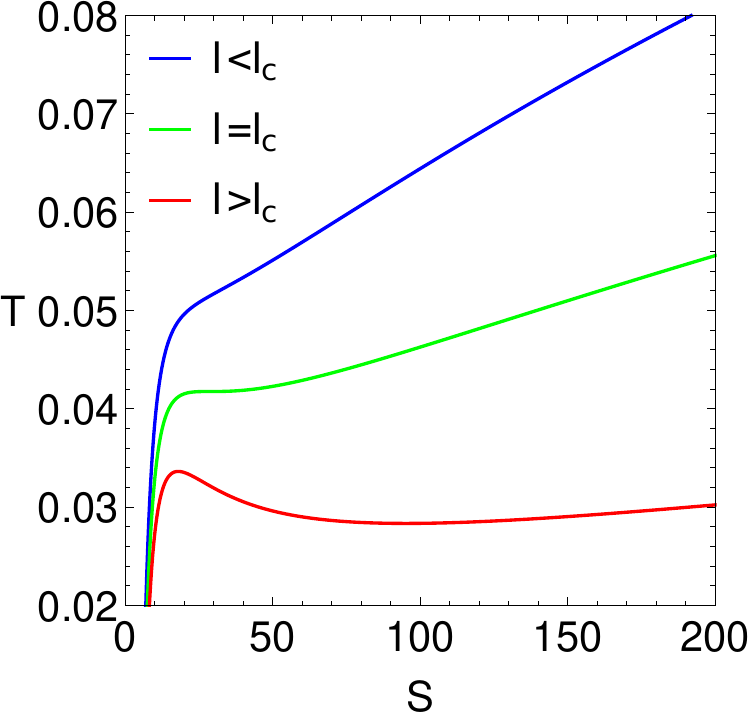}
  \end{subfigure}
  \vspace{0.5cm}
    \centering
  \begin{subfigure}{0.33\textwidth}
  \caption*{ $ J=0.3$}
    \includegraphics[width=\linewidth]{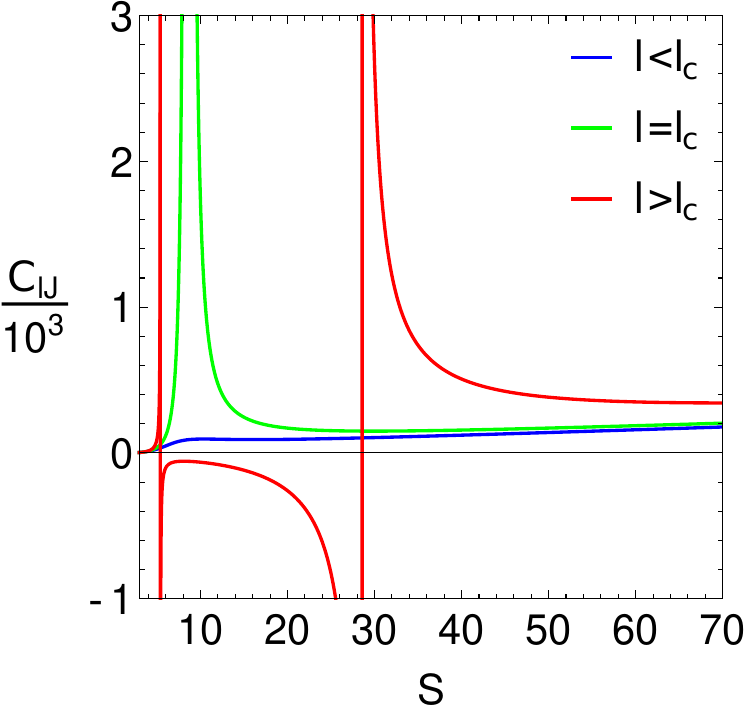}
  \end{subfigure}
  \hfill
  \begin{subfigure}{0.32\textwidth}
      \caption*{ $ J=0.6$}
    \includegraphics[width=\linewidth]{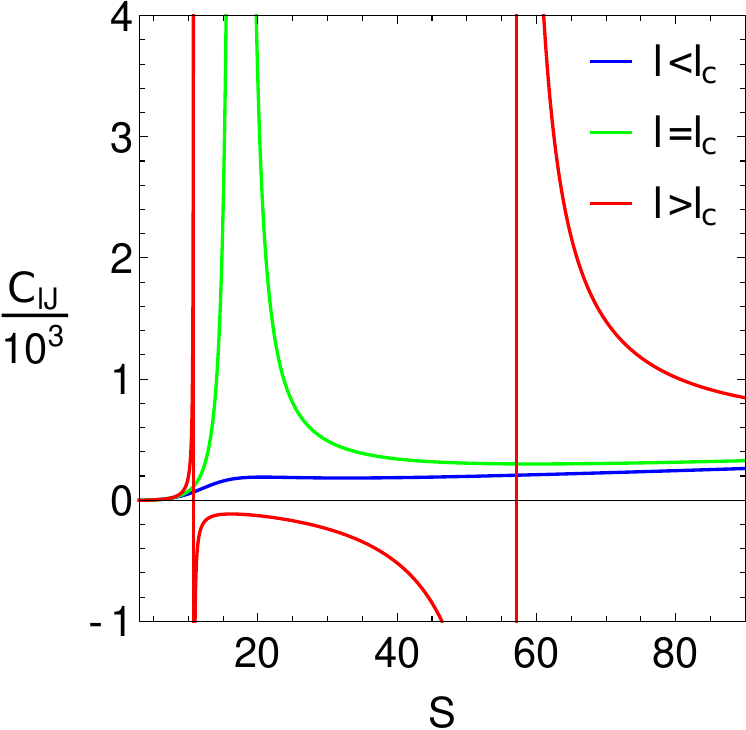}
  \end{subfigure}
  \hfill
  \begin{subfigure}{0.33\textwidth}
      \caption*{ $ J= 1$}
    \includegraphics[width=\linewidth]{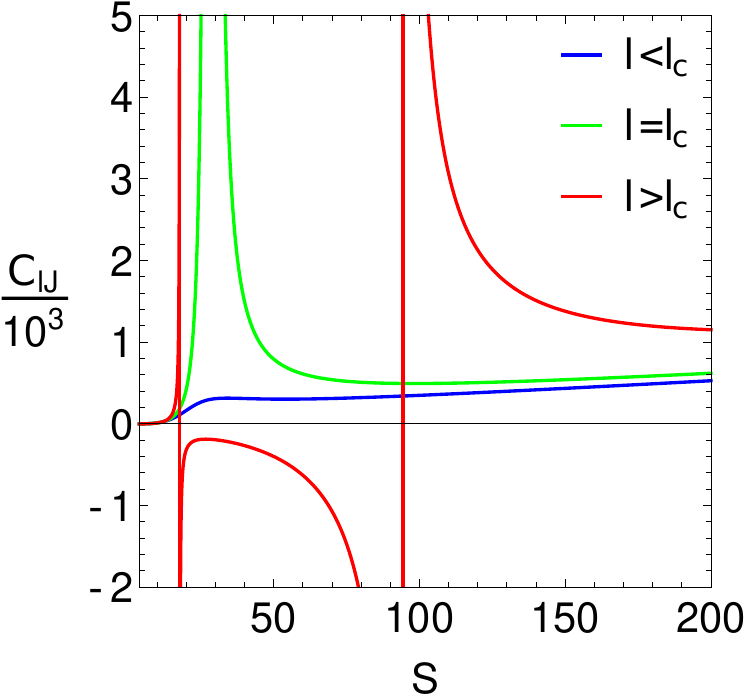}
  \end{subfigure}
  \caption{Temperature $T$ (top) and heat capacity $C_{lJ}$ (bottom) as functions of entropy $S$ for various values of $J$ and $l$. For \( l =l_c \), the local extrema merge into an inflection point. For smaller \( l \), the curve becomes monotonically increasing.}
  \label{fig:figura11_completa}
\end{figure}

In Fig.~\ref{temp_j_zero}, we illustrate  the temperature and heat capacity for \( J = 0 \), displaying the typical Schwarzschild-AdS behavior \cite{Ladino:2024ned}. The temperature curve initially decreases to a minimum, corresponding to the branch of small, unstable black holes (\( C_{lJ} < 0 \)), and then increases along the branch of large, stable black holes (\( C_{lJ} > 0 \)).  For \( J \neq 0 \) and \( l > l_c \), a stable small black hole branch appears, separated from large black holes by an intermediate unstable phase (Fig.~\ref{fig:figura11_completa}). This leads to a first-order phase transition between small (liquid) and large (gaseous) black holes, similar to the RN-AdS case \cite{Ladino:2024ned}. In extended thermodynamics, the Maxwell equal area law replaces the oscillatory region in the \( T-S \) diagram with an isotherm at the coexistence temperature \cite{kubizvnak2017black, spa}. Furthermore, as  \( l \) decreases, the local extrema of \( T(S, l, J) \) merge at \( l_c \), forming an inflection point and marking a second-order phase transition. For \( l < l_c \), the temperature increases monotonically, and the black hole enters a supercritical phase where small and large phases become indistinguishable \cite{wang2022ruppeiner}.   \\

\begin{figure}[H]
  \centering
  \begin{subfigure}{0.46\textwidth}
  \caption*{(a) $\hspace{6.2cm}$ $\vspace{0.2cm}$ }
    \includegraphics[width=\linewidth]{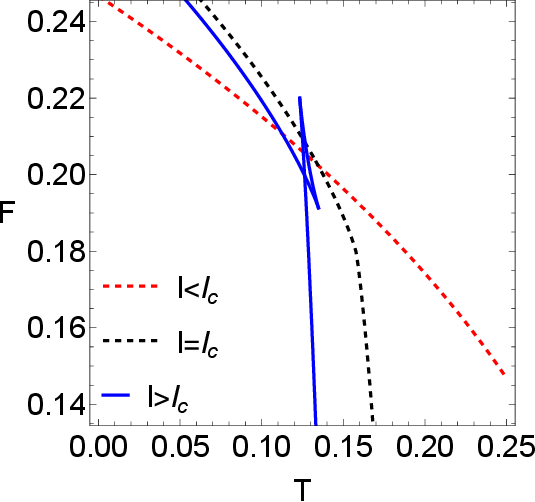}
  \end{subfigure}
  \hfill
  \begin{subfigure}{0.46\textwidth}
  \caption*{(b) $\hspace{6.2cm}$ $\vspace{0.2cm}$}
    \includegraphics[width=\linewidth]{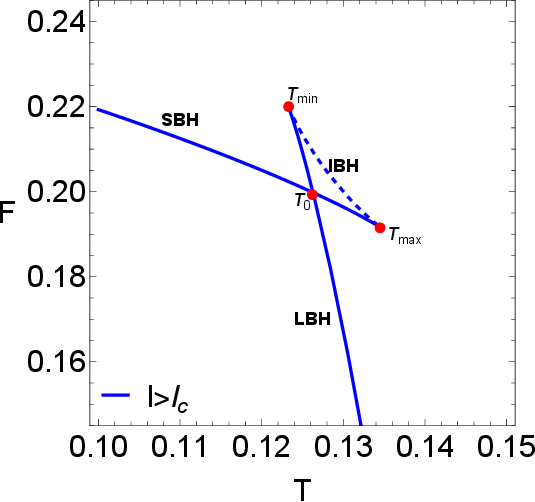}
  \end{subfigure}
  \caption{Free energy of the Kerr-AdS black hole in the canonical ensemble for fixed  $J=0.07$, $l=1.3 l_c$ (blue curve), $l=0.5 l_c$ (red curve). (a) The figure shows that for large curvature radius (i.e., above $l > l_c$), there are three branches of black hole solutions,
while for small $l$ (i.e., below $l < l_c$), there is only one branch of solutions with no special behavior. (b) Here we
set $l=1.3 l_c$. “SBH”, “IBH”, and “LBH” represent the small, intermediate
and large black holes, respectively.\label{gibbs_canonical}}
\end{figure}

In the canonical ensemble, the thermodynamic properties of the black hole are characterized by the Helmholtz free energy, defined as \cite{caldarelli2000thermodynamics}
\begin{align}
F(T,l,J) \equiv M-TS=\frac{1}{8M} \left[ J^2 \left( \frac{8}{l^2} + \frac{12\pi}{S} \right) + \frac{S \left(\pi^2 - \frac{S^2}{l^4}\right)}{\pi^3} \right],
\end{align}
where from Eq.~ (\ref{temp}) it is clear that $S$ is an implicit function of the thermodynamic variables. Moreover, using Eq.~ (\ref{temp}), we can find an analytic solution for $r_h(T,l,J)$. In particular, we observe that for $l>l_c$, in principle, there can coexist three branches of black hole solutions, as depicted in Fig.~ \ref{gibbs_canonical}.
In Fig.~\ref{gibbs_canonical}, the Kerr-AdS phase diagram matches that of RN-AdS black holes \cite{Ladino:2024ned}. The SBH/IBH meet at the inflection point \( T_{\max} \) and IBH/LBH at \( T_{\min} \). Thus, the LBH is the preferred state (lower free energy) when $T \in [T_0, T_{max}]$, and for $T < T_0$, the preferred state is the stable SBH. The \( F-T \) diagram shows a characteristic “swallowtail”, where the free energy of the black hole intersects with itself at $T = T_0$, indicating a first-order phase transition between SBH and LBH. The IBH branch, having a negative slope in \( T-S \), is thermodynamically unstable. This behavior is analogous to the gas-liquid phase transition in van der Waals fluids \cite{callen1998thermodynamics}.


\subsection{Grand-canonical ensemble}
\label{grand-cano}
In the grand canonical ensemble, where the angular velocity \( \Omega \) and curvature radius \( l \) are fixed, the phase structure is analyzed through the Gibbs free energy
\begin{align}
     G(T,l,\Omega)&\equiv M-TS-\Omega J=\frac{1}{8 \pi^3 l^4 M}\Bigg[\frac{S \left(l^2 \pi + S\right) \left(l^4 \pi^2 - S^2 + l^2 S^2 \Omega^2\right)}
{S + l^2 \left(\pi - S \Omega^2\right) }\Bigg].
\end{align}
Additionally, in this representation, the heat capacity reads
\begin{equation}
    C_{\Omega l}\equiv T\Big(\frac{\partial S}{\partial T}\Big)_{\Omega l}=\frac{2 S \big[l^2 (S \Omega^2-\pi )-S \big] \big[l^2 S (3 S \Omega^2-4 \pi )-3 S^2 - l^4 \pi (\pi - 2 S \Omega^2) \big] ( l^2 \pi + S)}
    { l^4 S^3 (4 l^2 \pi + 3 S) \Omega^4- 6 l^2 S^2 (l^2 \pi + S)^2 \Omega^2- \big[ (l^2 \pi - 3 S) (l^2 \pi + S)^3 \big]}.
\end{equation}
Expressing the Hawking temperature Eq.~(\ref{tempkerr}) in terms of $\Omega$, we obtain
\begin{align}
T(S,l,\Omega)=\sqrt{\frac{1}{S + l^2 \left(\pi - S \Omega^2\right)}}\left[
\frac{3 S^2 + l^2 S \left(4 \pi - 3 S \Omega^2\right) + 
    l^4 \pi \left(\pi - 2 S \Omega^2\right) }{4 \pi^{3/2} l^2  \sqrt{S \left(l^2 \pi + S\right)}}\right].
    \label{Temp_grand_canonical}
\end{align}
From the above equation, it follows that $T$ is well-defined only for values satisfying  
\begin{equation} 
\Omega < \Omega_{extremal}=\sqrt{\frac{(l^2 \pi + S) (l^2 \pi + 3 S)}{S (2 l^2 \pi + 3 S)}}, \label{wextremal}  
\end{equation}  
which ensures that $T$ remains a positive real function. The upper bound corresponds to the extremal black hole limit, where $T = 0$.  
This establishes a restriction on the angular velocity \( \Omega \) for a given entropy and curvature radius. However, a different bound for the angular velocity arises from holographic arguments \cite{caldarelli2000thermodynamics}. In this context, black holes with an angular velocity satisfying $\Omega l < 1$ correspond to thermal states in the dual field theory. Otherwise, the rotating Einstein universe, where the Conformal Field Theory (CFT) resides, would exceed the speed of light \cite{hawking1999rotation}. This condition is closely related to the absence of superradiant modes when $\Omega l < 1$ \cite{hawking2000stability}, as well as to the fact that for $\Omega l \geq 1$, no globally timelike Killing vector field can be defined outside the event horizon \cite{hawking1999rotation}.\\

In the grand-canonical ensemble, the nature of the small/large black hole phase transition differs from that in the canonical ensemble. In the latter, the heat capacity $C_{lJ}$ can exhibit up to two physical singularities (see Fig.~ \ref{fig:figura11_completa}), indicating the presence of an intermediate branch of unstable black holes. In the context of extended black hole thermodynamics \cite{kubizvnak2017black}, this phenomenon, further illustrated in Fig.~ \ref{gibbs_canonical}, can be interpreted either as a region of phase mixing \cite{kubizvnak2017black,spallucci2013maxwells} (first-order phase transition) or as the presence of two second-order phase transitions \cite{Ladino:2024ned}. In contrast, this feature is absent in the grand-canonical ensemble, where the heat capacity $C_{\Omega l}$ can exhibit at most one singularity. This is because the equation of state $T(S, l, \Omega)$ possesses only a local minimum and lacks an inflection point. This is not surprising, since it is well-known that the stability properties of a black hole can depend on the statistical ensemble \cite{caldarelli2000thermodynamics, gibbons1977action}. Interestingly, the phase transition structure of the Kerr-AdS black hole closely resembles that of the RN-AdS and Schwarzschild-AdS black holes~\cite{Ladino:2024ned}. A plausible explanation for this similarity is that all black holes in electrovacuum Einstein gravity belong to the same universality class
\cite{kubizvnak2017black}. This observation motivates further investigation into more general setups, such as black holes with Lifshitz scaling \cite{herrera2023anisotropic, romero2024extended1} and configurations featuring a hyperscaling-violating parameter \cite{herrera2023anisotropic, herrera2021hyperscaling}. \\

\begin{figure}[H]
  \centering
  \begin{subfigure}{0.44\textwidth}
  \caption*{(a) \hspace{7.7cm} \vspace{0.6cm}   }
    \vspace{-0.3cm} \includegraphics[width=\linewidth]{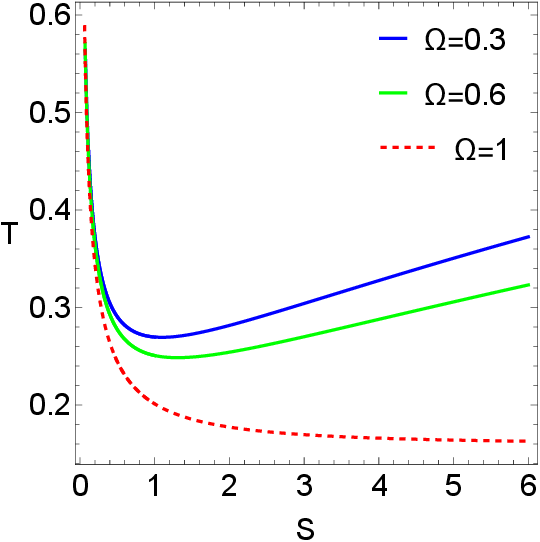}
  \end{subfigure}
  \hfill
  \begin{subfigure}{0.48\textwidth}
  \caption*{(b)  \hspace{7.5cm} \vspace{0.2cm}   } \includegraphics[width=\linewidth]{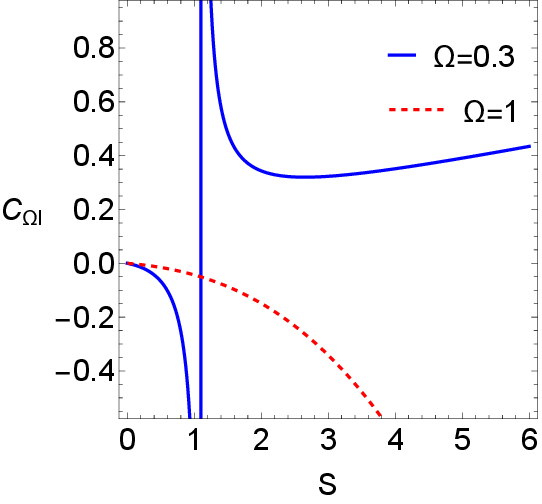}
  \end{subfigure}
  \vspace{0.6cm}
    \centering
  \begin{subfigure}{0.45\textwidth}
  \caption*{(c)  \hspace{7.5cm} \vspace{0.1cm}  }
   \hspace{-0.4cm}   \includegraphics[width=\linewidth]{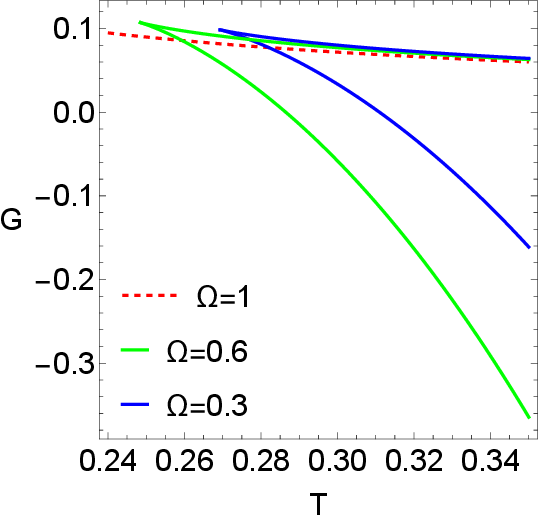}
  \end{subfigure}
  \hfill  
  \begin{subfigure}{0.475\textwidth}
  \caption*{(d)  \hspace{7.9cm} \vspace{0.1cm}  }
 \hspace{-0.1cm} \vspace{-0.15cm}   \includegraphics[width=\linewidth]{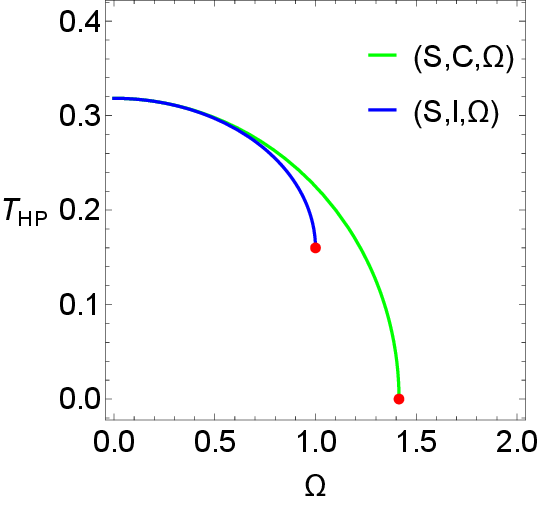}
  \end{subfigure}
  \caption{(a) Temperature \( T(S, l, \Omega) \), (b) Heat capacity \( C_{\Omega l}(S, l, \Omega) \), (c) Gibbs free energy \( G(T, l, \Omega) \), and (d) Hawking-Page temperatures \( T_{HP}(S, C, \Omega) \) and \( T_{HP}(S, l, \Omega) \), for \( l = 1 \) and different values of angular velocity \( \Omega \). The red dotted points indicate the minimum values of \( T_{HP} \). }
  \label{grandcanonical}
\end{figure}

As shown in Fig.~ \ref{grandcanonical}, we analyze the stability of Kerr-AdS black hole phase transitions in the grand-canonical ensemble. In Fig.~ \ref{grandcanonical}(a), we present the equation of state \( T(S, l, \Omega) \) for a fixed \( l = 1 \) and varying angular velocity \( \Omega \). When \( \Omega = 1 \), the temperature asymptotically approaches \( T \to 1/2\pi \) as \( S \to \infty \), with no inflection point observed. However, for \( \Omega < 1 \), a local minimum emerges, indicating a phase transition between small and large black holes. Fig.~ \ref{grandcanonical}(b) further illustrates the stability of black hole phases and the location of the phase transition through the behavior of the heat capacity \( C_{\Omega l} \). Meanwhile, Fig.~ \ref{grandcanonical}(c) depicts the Gibbs free energy for a fixed \( l = 1 \) and varying \( \Omega \). When \( \Omega = 1 \), only a single, trivial black hole branch exists. In contrast, for \( \Omega < 1 \), two distinct branches appear: the upper branch corresponds to thermally unstable small black holes (\( C_{\Omega l} < 0 \)), while the lower branch represents thermally stable large black holes (\( C_{\Omega l} > 0 \)). \\

Furthermore, the above analysis  concerns only local thermodynamic stability and is not sufficient to ensure global stability (for a review of global stability in different gravity models, see \cite{avramov2024thermodynamic,avramov2024thermodynamic2}). In principle, there may exist a reference background, such as thermal AdS space, with lower free energy, which would then become the preferred thermodynamic state. This phenomenon is known as the Hawking-Page (HP) phase transition \cite{hawking1983thermodynamics}. In the context of the AdS/CFT
correspondence \cite{maldacena1999large}, this transition, was later understood
as a confinement/deconfinement phase transition in the boundary CFT \cite{witten1998anti2}. It is important to note that AdS space itself does not possess angular momentum, making it an unsuitable reference background for the Kerr-AdS solution in the canonical ensemble, where $J$ is fixed \cite{caldarelli2000thermodynamics}. Instead, the free energy of the black hole should be compared with that of alternative solutions, such as AdS space filled with rotating thermal radiation \cite{chamblin1999holography}. Consequently, the precise nature of the HP phase transition and its holographic interpretation in the rotating case remains an open question and warrants further investigation. Nonetheless, one can expect a
phase transition between the thermal gas, and the locally stable large black hole state. It is generally assumed that $G=0$ for a thermal gas, so the precise temperature at which the phase transition takes place can be determined by calculating the non-trivial radius of the event horizon for which the black hole solution exhibits vanishing Gibbs free energy. Substituting this result into Eq.~ (\ref{Temp_grand_canonical}), we obtain
\begin{equation}
T_{HP}=\frac{1+\sqrt{1-(\Omega l)^2}}{2 \pi l}.
\end{equation}
When $\Omega \to 0$, the HP temperature for a Schwarzschild-AdS black hole is recovered, with $T_{HP} \to 1/\pi l$ \cite{kubizvnak2017black,Ladino:2024ned}. In \cite{gao2022thermodynamics}, a distinct $T_{HP}$ was derived using the formalism of restricted phase space thermodynamics (RPST) \cite{gao2022restricted} in the slow-rotation limit. Similarly, in \cite{belhaj2020universal}, which employed the standard black hole thermodynamic approach, we found a discrepancy in the value of $T_{HP}$ despite agreeing on all the thermodynamic parameters. 

As shown in Fig.~ \ref{grandcanonical}(d), the Hawking-Page temperature behavior for the Kerr-AdS black hole is presented for a fixed curvature radius \( l = 1 \) and varying angular velocity \( \Omega \). The blue curve corresponds to the case where \( l \) is the thermodynamic parameter, while the green curve represents the scenario where the central charge \( C = l^2 / G \) is used instead, resulting in a different behavior in the slow-rotation regime \cite{gao2022thermodynamics}. The red dotted points indicate the minimum values of \( T_{HP} \). In Fig.~\ref{grandcanonical}(d), a difference in the values of the Hawking-Page temperature is observed. This discrepancy may arise from the RPST framework, which treats the holographic central charge as a thermodynamic parameter, thereby defining the fundamental thermodynamic relation, Eq.~\eqref{fundame1}, as a homogeneous function of degree one. In contrast, our approach considers the curvature radius \( l \) as a thermodynamic variable, leading to a quasi-homogeneous fundamental relation, as discussed in Sec.~\ref{GTDsection}. Moreover, using Eq.~\eqref{wextremal} and the restriction \( a < l \), along with the extremality condition given by Eq.~\eqref{extremalcondition}, it can be shown that extremal Kerr-AdS black holes do not admit a Hawking-Page temperature, similar to the extremal RN-AdS case \cite{Ladino:2024ned}. However, an erroneous result may arise from the slow-rotation approximation \cite{gao2022thermodynamics}, which incorrectly predicts the existence of a Hawking-Page temperature for extremal black holes.  Additionally, as argued earlier, there appears to be a universality class for AdS electrovacuum black hole solutions \cite{kubizvnak2017black}.  
For spherical Schwarzschild-AdS and RN-AdS black holes, it has been demonstrated that the ratio\footnote{$T_{min}$ corresponds to the local minimum of Eq.~ (\ref{Temp_grand_canonical}), which is too extensive to be fully written here.}  $\zeta=T_{HP}/T_{min}$, is a universal constant \cite{wei2020novel}, which, for the 4-dimensional case, is given by $2/\sqrt{3}$ \cite{belhaj2020universal,Ladino:2024ned}. The determination of such scalar constants is crucial for understanding black hole microstructure and for achieving a precise quantitative description of fundamental theories of the universe. A closer inspection reveals that $\zeta$ depends on the spacetime dimension and/or the symmetries of the theory \cite{wei2020novel}. For instance, when black holes exhibit Lifshitz scaling, this ratio deviates from the standard AdS value \cite{romero2024extended1}. Unfortunately, for the Kerr-AdS black hole, an analytical expression for $\zeta$ cannot be obtained. However, a Taylor expansion for $\Omega l \ll 1$ yields  
\begin{equation}
\zeta = \frac{2}{\sqrt{3}} + \frac{9 \sqrt{3}}{128 \times 2^{1/3}} (\Omega l)^{8/3} + \mathcal{O}\big(\Omega l\big)^{10/3}. \label{ratio1}
\end{equation}
In the limit \( \Omega \to 0 \), the universal ratio for 4-dimensional non-rotating AdS black holes is recovered. Notably, our findings differ from those in~\cite{belhaj2020universal}. Within the RPST framework, a result of \( \zeta = 2/\sqrt{3} \) for the Kerr-AdS solution in the slow-rotation limit was obtained in~\cite{gao2022thermodynamics}, which can be interpreted as the leading term of Eq.~\eqref{ratio1}.  

\section{Geometrothermodynamics and Microstructure}
\label{GTDsection}
One of the main ingredients of GTD is Legendre invariance, which in ordinary equilibrium thermodynamics means that the properties of a system do not depend on the choice of thermodynamic potential used for its description. GTD incorporates Legendre invariance into the formalism by introducing the auxiliary phase space ${\cal T}$, which is a $2n+1$ dimensional manifold with metric $G_{AB}$, $A,B=0,1,...,2n$, where $n$ is the number of thermodynamic degrees of freedom. For concreteness, we introduce the set of coordinates $Z^A = \{ \Phi, E^a, I_a\}$ with $a=1,...,n$. Then, the line elements $G =G_{AB}dZ^AdZ^B$  of the Legendre invariant metrics of ${\cal T}$ can be expressed as 
\begin{equation}
    G^{I/II}=\left(d\Phi-I_adE^a\right)^2+(\xi_{ab}E^aI^b)(\chi_{cd}dE^cdI^d),
\end{equation}
\begin{equation}
    G^{III}=\left(d\Phi-I_adE^a\right)^2+\sum_{a=1}^{n}\xi_{a}(E_aI_a)^{2k+1}dE^adI^a,
\end{equation}
where $\delta^c_a=diag(1,1, \ldots , 1)$, $\eta^c_a=diag(-1, \ldots , 1)$, and $\xi_{a}$ are real constants. Furthermore,  $\chi_{cd} = \delta_{cd}$ for $G^I$ and $\chi_{cd} = \eta_{cd}$ for $G^{II}$, $\xi_{ab}$ is a diagonal ($n\times n$) real matrix, and $k$ is an integer. The corresponding induced metrics on the equilibrium space are:
\begin{equation}
    g^I_{ab}=\beta_\Phi \Phi \delta^c_a\frac{\partial^2 \Phi}{\partial E^b \partial E^c}, \label{g1}
\end{equation}
\begin{equation}
    g^{II}_{ab}=\beta_\Phi \Phi \eta^c_a\frac{\partial^2 \Phi}{\partial E^b \partial E^c}, \label{g2}
\end{equation}
\begin{equation}
    g^{III}=\sum_{a=1}^{n}\beta_a\left(\delta_{ad}E^d\frac{\partial \Phi}{\partial E^a}\right)\delta^{ab}\frac{\partial^2 \Phi}{\partial E^b \partial E^c} dE^adE^c.\label{g3}
\end{equation}
To obtain the components of the metrics $g^I$ and $g^{II}$, we have chosen $\xi_{ab}=\beta_{ab}={\rm diag}(\beta_1,\beta_2,...,\beta_n)$ and $\xi_a=\beta_a$, where $\beta_{a}$ are the quasi-homogeneous coefficients of the fundamental equation $\Phi=\Phi(E^a)$, and used the quasi-homogeneous Euler identity, which generates the conformal factor $\beta_\Phi \Phi$. Moreover, to guarantee that the line element $g^{III}$ leads to results compatible with those of $g^I$ and $g^{II}$, we set $k=0$.

\begin{figure}[H]
\begin{minipage}[t]{0.32\linewidth}
 \centering
\hspace{1cm}
\includegraphics[width=1\linewidth]{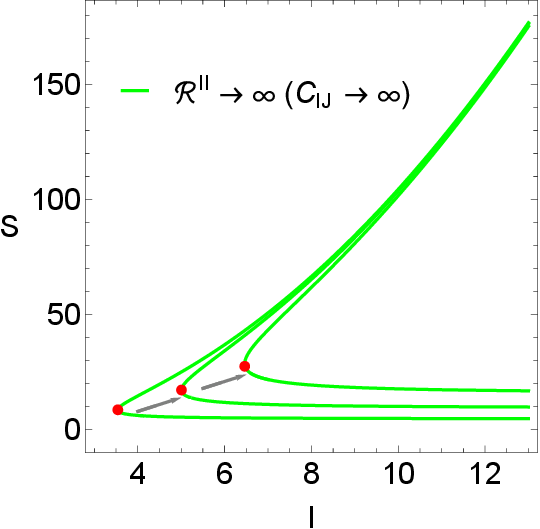}
 (a)\hspace{10cm}
\end{minipage}%
\hfill%
\begin{minipage}[t]{0.32\linewidth}
 \centering
\hspace{1cm}
\includegraphics[width=1\linewidth]{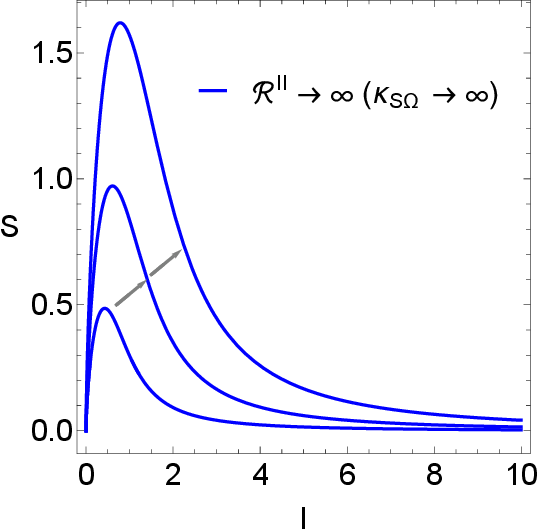}
(b)\hspace{12cm}
\end{minipage}%
\hfill%
\begin{minipage}[t]{0.32\linewidth}
 \centering
\hspace{1cm}
\includegraphics[width=1\linewidth]{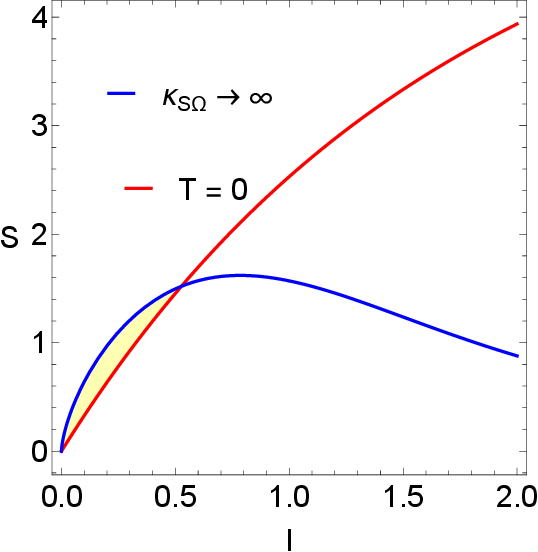}
(c)\hspace{12cm}
\end{minipage}%
\hfill%
\caption{Critical curves of \( \mathcal{R}^{II} \) for \( J = 0.3, 0.6, 1 \) (arrows indicate increasing \( J \)).  
(a) Along the green curves, \( \mathcal{R}^{II} \) diverges with \( C_{lJ} \), indicating a second-order phase transition. Red dots mark the critical points of \( T(S, l, J) \), given by \( l_c \).  
(b) The blue curves represent divergences in \( \mathcal{R}^{II} \) caused by divergences in $\kappa_{S\Omega}$.  
(c) The red curve represents the extremal black hole curve for \( J=1 \). The divergence associated to 
$\kappa_{S\Omega}$ occurs with \( T > 0 \) only within the yellow-shaded region.}
\label{jcritical}
\end{figure}

For Kerr-AdS black holes, the most general scenario involves considering a three-dimensional thermodynamic equilibrium space. While it is possible to work with a reduced two-dimensional equilibrium space, such an approach may yield inconsistent results, as it neglects fluctuations in specific thermodynamic parameters (see, for instance, \cite{mirza2007ruppeiner}). Performing the rescaling of the extensive variables, it is easy to see that Eq.~ (\ref{fundame1}) is a quasi-homogeneous function of degree $\beta_M$, 
if the condition
\begin{equation}
    \beta_l=\frac{1}{2}\beta_S,\quad \beta_J=\beta_S, \quad \beta_S\equiv 2\beta_M, \label{escal2}
\end{equation}
are imposed \cite{quevedo2019quasi}. With this condition, it is then trivial to check that the  Euler identity 
\begin{equation}
M = 2TS + 2\Omega J + \Psi l
\end{equation}
is fulfilled. The equation above is simply the well-known Smarr relation for rotating AdS black holes \cite{caldarelli2000thermodynamics}. This relation arises directly from the quasi-homogeneous nature of the black hole solution. Therefore, we can use the results presented in \cite{Ladino:2024ned}, which use explicitly the Euler identity.
According to Eqs.~ (\ref{g1})-(\ref{g3}), the line elements\footnote{Notation: $M_{,xy}\equiv \partial^2 M/\partial x \partial y$.} of the GTD metrics of the Kerr-AdS black hole can be written as 
\begin{align}
    g^I&= \beta_M M\Big(M_{,SS}
 dS^2+2M_{,SJ}dSdJ+2M_{,Sl}dS dl+2M_{,Jl}dJdl+M_{,JJ}dJ^2
+ M_{,ll}dl^2\Big),\\
 g^{II} &= \beta_M M\Big(-M_{,SS}
 dS ^2+2M_{,Jl}dJdl+M_{,JJ}dJ^2
+ M_{,ll}dl^2\Big),\label{metric2}\\ 
    g^{III}&=\beta_S \big(T S \big)M_{,SS}dS^2+\beta_J \big(\Omega J \big)M_{,JJ}dJ^2 +\beta_l \big(\Psi  l \big)M_{,ll}dl^2+M_{,SJ}\Big[\beta_S \big(TS \big)+\beta_J \big(\Omega J \big)\Big]dSdJ+ \notag\\
    &M_{,Sl}\Big[\beta_S \big(TS\big)+\beta_l\big(\Psi  l \big)\Big]dSdl+M_{,Jl}\Big[\beta_J \big(\Omega J  \big)+\beta_l \big(\Psi l \big)\Big]dJdl. 
\end{align}
The explicit expressions for the second derivatives of the thermodynamic potential are given in Eq.~ \eqref{second_deriv} of Appendix \ref{apenA}. The general structure of the corresponding independent curvature scalars has been analyzed in \cite{Ladino:2024ned}, where we obtained the general conditions that relate the singularities of the three GTD metrics, which in this case can be written as 
\begin{align}
I&: M_{,SS}\Big [ \big (M_{,Jl}\big)^2-M_{,JJ}M_{,ll} \Big]+M_{,JJ}\big (M_{,Sl}\big)^2+M_{,ll}\big (M_{,SJ}\big)^2-2 M_{,SJ} M_{,Sl} M_{,Jl} =0 \label{RNcondI}
    \\
     II&: M_{,SS}\Big[(M_{,Jl}\big)^2-M_{,JJ}M_{,ll}\Big]=0,\\
     III&: M_{,SS}=M_{,SJ}=M_{,Sl}=0, \quad \text{or}  \quad M_{,SJ}=M_{,JJ}=0,\nonumber \\ &\text{or} \quad M_{,Sl}=M_{,ll}=0,\quad \text{or} \quad M_{,JJ}=M_{,ll}=0. \label{RNcondIII}
\end{align}
Furthermore, we align the above singularity conditions, with the system’s  thermodynamic response functions (Appendix \ref{apenA}) to obtain
\begin{align}
    I&: \frac{T}{C_{ \Psi\Omega  }\kappa_{S\Omega}\kappa_{Sl}}=0 ,\label{uy2}\\
    II&: \frac{T }{C_{lJ}\kappa_{S\Omega}\kappa_{Sl}}=0,\label{qa}\\
    III&: \frac{1}{C_{lJ}}=\frac{1}{\alpha_{Sl}}=\frac{1}{\alpha_{SJ}}=0,\quad \text{or} \quad
    \frac{1}{\alpha_{Sl}}=\frac{1}{\kappa_{Sl}}=0,\label{rre3}\\ \nonumber
    &\text{or} \quad \frac{1}{\alpha_{S J}}=\frac{1}{\kappa_{SJ}}=0,\quad\text{or} \quad  \frac{1}{\kappa_{S J}}=\frac{1}{\kappa_{Sl}}=0.
\end{align}
From Eqs.~ \eqref{uy2}-\eqref{rre3}, we can conclude that there exists a one-to-one correspondence between the divergences of the scalar curvature of the GTD with the divergences of the response functions of the system. Additionally, using Eqs.~ (\ref{second_deriv}) it is straightforward to check  that the singularity conditions $I$ and $III$ cannot generally be satisfied for values of $S > 0$ and $l > 0$. The singularity condition $II$ aligns with the critical curves of the heat capacity $C_{lJ}$ and/or the compressibility $\kappa_{S\Omega}$ (see Fig.~ \ref{jcritical}), since $\kappa_{Sl}$ remains regular. 

Solving condition $II$, we find the critical surfaces where the scalar curvature $\mathcal{R}^{II}$ diverges, which are given by
\begin{align}
& J_{critical1}^2= \frac{
    -3 S^2 (l^2 \pi + S)^2 (l^2 \pi + 2 S) \pm 
    2 \sqrt{S^4 (l^2 \pi + S)^3 (3 l^6 \pi^3 + 10 l^4 \pi^2 S + 
    15 l^2 \pi S^2 + 9 S^3)} 
    } 
    {4\pi^4(3 l^6 \pi + 4 l^4 S)}, \label{con1}\\
& J_{critical2}^2=\frac{3 l^2 S(\pi l^2+S)^2 \pm \sqrt{3} \sqrt{l^4 S^2 (l^2 \pi S)^3 (3 l^2 \pi + 11 S)}}{8 l^4 \pi^3}\label{con2}.
\end{align}
For different combinations of \( S \) and \( l \), each \( J \)-function exhibits a double root, leading to four values of \( J \) at which the scalar curvature of \( g^{II} \) diverges. However, for \( l > 0 \) and \( S > 0 \), \( J^2 \) has only one physically acceptable solution. When working in the canonical ensemble, it is convenient to express Eqs.~ \eqref{con1}–\eqref{con2} as critical surfaces for the entropy, as depicted in  Fig.~ \ref{jcritical}.
\begin{figure}[H]
\begin{minipage}[t]{0.32\linewidth}
 \centering
\hspace{1cm}
\includegraphics[width=1\linewidth]{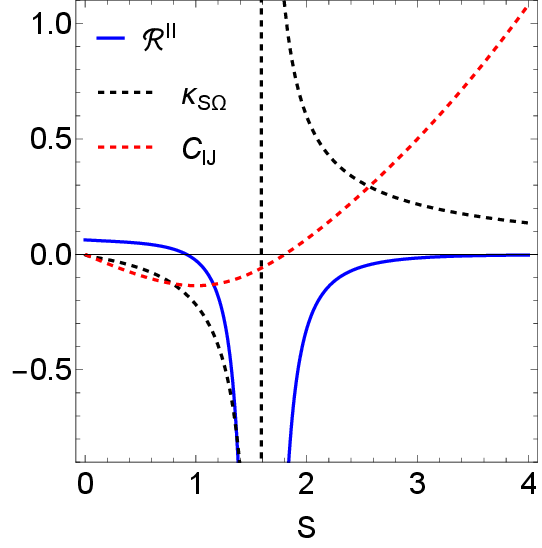}
 (a)\hspace{10cm}
\end{minipage}%
\hfill%
\begin{minipage}[t]{0.32\linewidth}
 \centering
\hspace{1cm}
\includegraphics[width=1\linewidth]{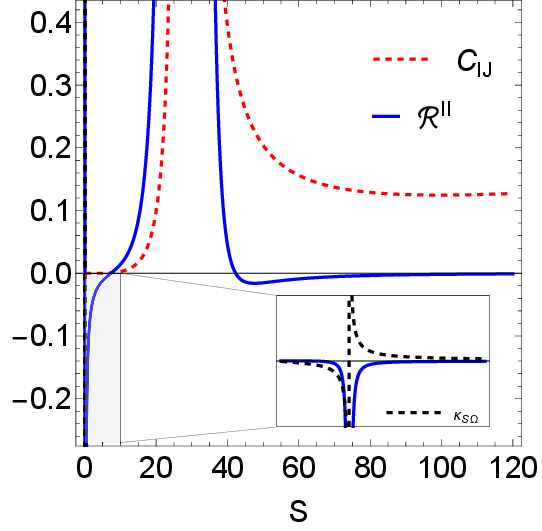}
(b)\hspace{12cm}
\end{minipage}%
\hfill%
\begin{minipage}[t]{0.32\linewidth}
 \centering
\hspace{1cm}
\includegraphics[width=1\linewidth]{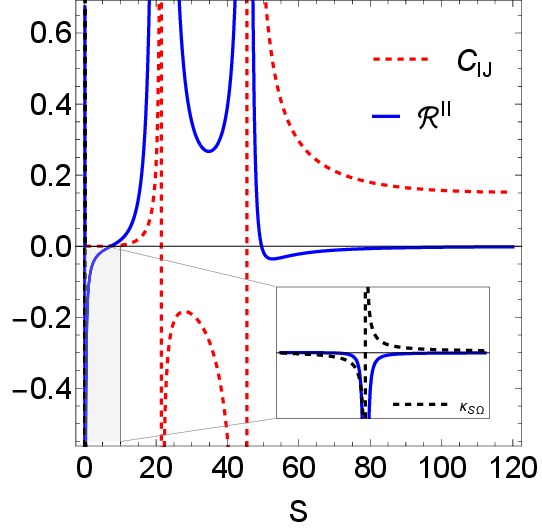}
(c)\hspace{12cm}
    \end{minipage}%
\hfill%
\caption{Normalized Ricci scalar $\mathcal{R}^{II}$ for different  radii of curvature $l$, and fixed parameters $J=1$, $\beta_M=1$. (a) For $l = 0.1l_c$,  (b) for $l = l_c$, and (c) for $l = 1.1l_c$.}

\label{RII}
\end{figure}
It is important to emphasize that the divergences in \( \kappa_{S\Omega} \) occur for black hole solutions with positive temperature only when the critical curve of Eq.~ \eqref{con2} lies above the extremal curve of Eq.~ \eqref{smintemperat}, corresponding to a small region around near-extremal black holes with small curvature radii \( l \), as shown in Fig.~ \ref{jcritical}. However, a closer inspection of the compressibility parameter \( \kappa_{S\Omega} \) reveals that its divergences occur for \( \Omega l \geq 1 \), a regime that can be excluded from the thermodynamic analysis, as discussed in Sec. \ref{grand-cano}. On the other hand, in Fig. \ref{RII}, \( \mathcal{R}^{II} \) as a function of entropy \( S \), it is evident that GTD accurately captures the Kerr-AdS black hole phase structure within the canonical ensemble. In Fig. \ref{RII}(a), for \( l < l_c \), \( \mathcal{R}^{II} \) exhibits a divergence corresponding to \( \kappa_{S\Omega} \). In Fig.~ \ref{RII}(b), for \( l = l_c \), \( \mathcal{R}^{II} \) shows two singularities: one corresponding to \( C_{lJ} \) and the other to \( \kappa_{S\Omega} \). In Fig.~ \ref{RII}(c), for \( l > l_c \), \( \mathcal{R}^{II} \) shows three singularities: two corresponding to \( C_{lJ} \) and one to \( \kappa_{S\Omega} \). However, the singularities corresponding to \( \kappa_{S\Omega} \) in Figs.~ \ref{RII}(b) and \ref{RII}(c) are  not  physical, as in these regions \( T < 0 \) and \( \Omega l > 1 \). This analysis suggests that the singularity in \( \kappa_{S\Omega} \) arises from considering \( l \) as a thermodynamic parameter. Consequently, extending the thermodynamic space could introduce additional singularities in the Kerr-AdS black hole. In contrast, using \( \Lambda = -3/l^2 \) as a state parameter avoids new singularities in Kerr-AdS black holes, as the term \( (M_{,J\Lambda})^2 - M_{,JJ}M_{,\Lambda\Lambda} \) remains positive for a negative cosmological constant \cite{larranaga2012geometric}.\\


In GTD, thermodynamic interaction corresponds to the curvature of the equilibrium manifold \cite{quevedo2007geometrothermodynamics}. The curvature scalar of GTD encodes the nature of the thermodynamic microscopic interaction: a positive scalar (\(\mathcal{R}^{II} > 0\)) indicates a repulsive interaction, a negative scalar (\(\mathcal{R}^{II} < 0\)) an attractive one, and zero curvature (\(\mathcal{R}^{II} = 0\)) corresponds to no interaction, as in an ideal gas, which is represented by a flat geometry \cite{ruppeiner1979thermodynamics}. \\

\begin{figure}[H]
  \centering
  \begin{subfigure}{0.49\textwidth}
  \caption*{(a) \hspace{7.5cm} \vspace{-0.5cm}   }
    \includegraphics[width=\linewidth]{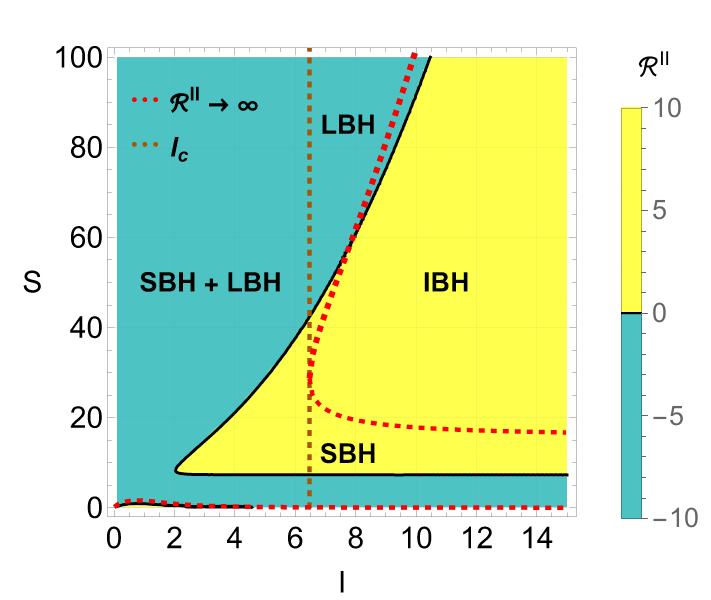}
  \end{subfigure}
  \hfill
  \begin{subfigure}{0.49\textwidth}
  \caption*{(b)  \hspace{7.5cm} \vspace{-0.5cm}   }
    \includegraphics[width=\linewidth]{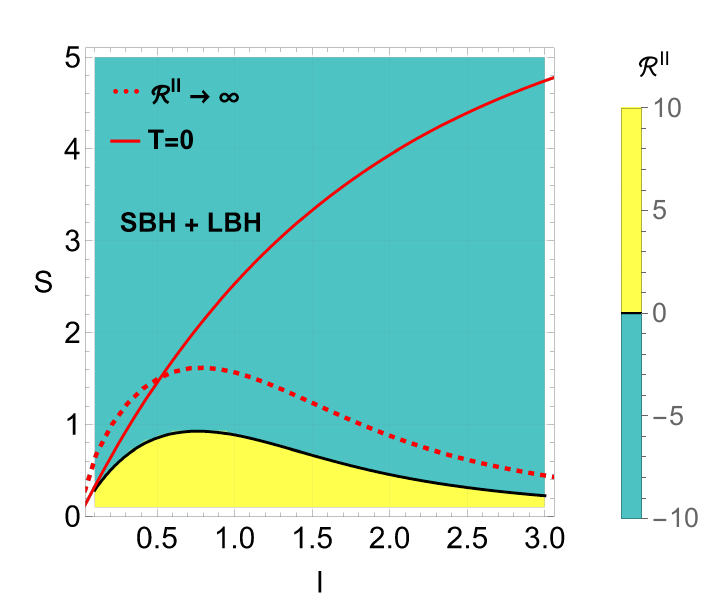}
  \end{subfigure}
  \vspace{0.5cm}
    \centering
  \begin{subfigure}{0.48\textwidth}
  \caption*{(c)  \hspace{7.3cm} \vspace{-0.5cm}  }
    \includegraphics[width=\linewidth]{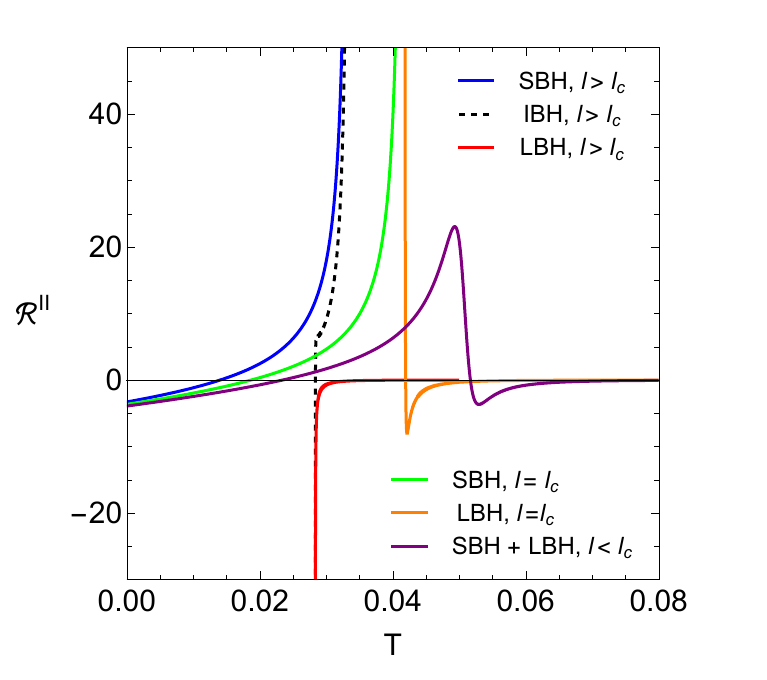}
  \end{subfigure}
  \hfill  
  \begin{subfigure}{0.465\textwidth}
  \caption*{(d)  \hspace{7.9cm} \vspace{-0.5cm}  }
 \hspace{-0.8cm}   \includegraphics[width=\linewidth]{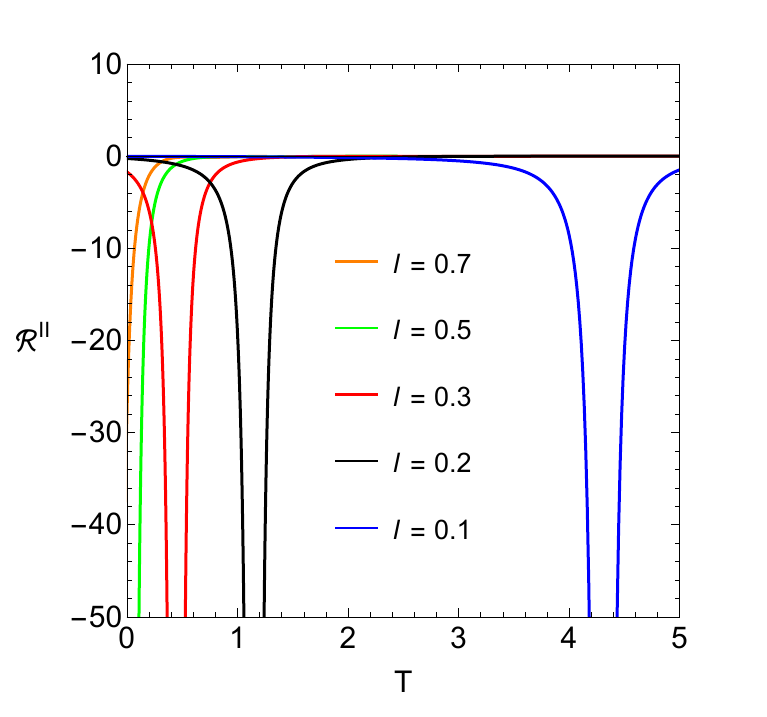}
  \end{subfigure}
  \caption{Microstructure behavior of Kerr-AdS black holes (\(J=\beta_M=1\)). (a) Contour map of \(\mathcal{R}^{II}\) highlighting regions where it is positive, negative, or zero. The two dashed red curves indicate the divergence of \(\mathcal{R}^{II}\), one above and one below. The figure is divided into four regions: to the left of \(l_c\), SBH and LBH are indistinguishable; to the right, the upper red dashed line separates the SBH (below), LBH (above), and IBH (intermediate) phases. (b) Contour map of \(\mathcal{R}^{II}\) emphasizing its divergences for small \(S\) and \(l\). In (c) and (d), \(\mathcal{R}^{II}\) is shown as a function of \(T\): (c) for different \(l\) values in each phase, and (d) highlighting its divergences for small \(S\) and \(l\).}
  \label{ff}
\end{figure}

The microstructure of Kerr-AdS black holes, as revealed by thermodynamic Ruppeiner geometry, displays behavior that contrasts with the results derived from the GTD formalism \cite{Hazarika, Banerjee, ref20, wei2021general, Zangeneh_2018, Dehyadegari, Sahay}.  In Fig.~ \ref{ff}, we show the microstructural behavior of Kerr-AdS black holes for \(J=\beta_M=1\). In Fig.~ \ref{ff}(a), we present the contour plot of \(\mathcal{R}^{II}\), highlighting regions where it is positive, negative, or zero. The two red dashed curves indicate the points where \(\mathcal{R}^{II}\) diverges, one above and the other below, corresponding to the singularities of the heat capacity \(C_{lJ}\) and the compressibility parameter \(\kappa_{S\Omega}\), respectively. The diagram is divided into four regions: to the left of \(l_c\), SBH and LBH are indistinguishable, while to the right of \(l_c\), the upper red dashed line, where \(\mathcal{R}^{II} \to \infty\), separates the SBH phase below, the LBH phase above, and the IBH phase in between. In Fig.~ \ref{ff}(b), we zoom into the contour plot of \(\mathcal{R}^{II}\), emphasizing its singularities for small values of \(S\) and \(l\), which correspond to the divergences of \(\kappa_{S\Omega}\). In Fig.~ \ref{ff}(c) and Fig.~ \ref{ff}(d), the scalar \( \mathcal{R}^{II} \) is shown as a function of temperature \( T \) for different values of \( l \): in Fig.~ \ref{ff}(c) for the typical behavior of the black hole phases, and in Fig.~ \ref{ff}(d) for small values of \( l \) and \( S \), highlighting the divergences corresponding to those of \( \kappa_{S\Omega} \).  \\

As shown in Fig.~\ref{ff}, Kerr-AdS black holes exhibit both attractive and repulsive thermodynamic interactions. For \(l \geq l_c\), at very low temperatures, only the SBH branch exists, whereas at higher temperatures, only the LBH branch remains. At intermediate temperatures, the curvature scalar exhibits two divergences, corresponding to phase transitions where the black hole transitions between the stable SBH/unstable IBH/stable LBH phases. In this regime, all three phases coexist. Moreover, the results indicate that thermodynamic interactions are stronger at intermediate temperatures, while they weaken for nearly extremal SBHs and superheated LBHs. As a result, a first-order phase transition is expected at intermediate temperatures, where the three branches coexist. The strong repulsive interactions in the SBH phase may drive the expansion of the event horizon, ultimately leading to a transition into the LBH phase. Additionally, when \(l = l_c\), the intermediate branch vanishes, turning the phase transition into a second-order one. For \(l < l_c\), the black hole enters a stable supercritical phase, where the same behavior is observed: attractive/repulsive/attractive interactions at low, intermediate, and high temperatures, respectively.\\\\
Furthermore, while superheated LBHs exhibit weak thermodynamic interactions and effectively behave like an ideal gas, a novel result from GTD, previously reported in \cite{Ladino:2024ned}, reveals that for \(l = l_c\), LBHs can exhibit strong repulsive interactions at intermediate temperatures, as illustrated in Fig.~\ref{ff}. On the other hand, some studies on Kerr-AdS black holes use the Ruppeiner thermodynamic approach, revealing a very different phase structure and microscopic interaction \cite{Hazarika, Banerjee, ref20, wei2021general, Zangeneh_2018, Dehyadegari, Sahay}. In general, these previous works show that the interaction in the SBH phase is purely attractive, or that there are configurations where it becomes repulsive at low temperatures and attractive at high temperatures. This contrasts with our results using the GTD formalism, where the interaction for SBH is attractive at low temperatures and repulsive at intermediate temperatures. Finally, within the context of the Ruppeiner approach, the interaction in the LBH or near-extremal black hole cases is typically dominated by a microstructure that is either strongly repulsive or strongly attractive, depending on the chosen thermodynamic variables. This contrasts with our results, which indicate weak interactions in these regimes. The variability in these results arises primarily from one of the key differences between the approaches: the choice of thermodynamic variables in the generalized thermodynamic space used to geometrically describe the system.\\
We observe that the phase structure and microstructure of Kerr-AdS black holes closely resemble those of RN-AdS black holes \cite{Ladino:2024ned}, suggesting an underlying universal behavior governing these gravitational systems. One of the few differences in their microscopic interactions, under the GTD formalism, is that the SBH phase of RN-AdS black holes is purely repulsive, whereas in Kerr-AdS black holes, it can involve both attractive and repulsive interactions. In other cases, however, they behave very similarly.\\

Within the Ruppeiner framework, some studies do not consider the cosmological constant \( \Lambda \) of the Kerr-AdS solution, or any quantity associated with the AdS background, as a thermodynamic variable (e.g., \cite{Hazarika, Banerjee, ref20}).  In contrast, other works adopt the extended thermodynamic approach, interpreting the cosmological constant as a pressure term, \( P = -\Lambda/8\pi = 3/8\pi l^2 \), and including it as an additional thermodynamic parameter, with its conjugate identified as a thermodynamic volume \cite{wei2021general, Zangeneh_2018, Dehyadegari, Sahay}. In some of these cases, only a reduced equilibrium space is considered, as seen in \cite{wei2021general, Zangeneh_2018, Dehyadegari}. However, this may lead to inconsistent results by neglecting fluctuations in certain thermodynamic parameters (see, for instance, \cite{mirza2007ruppeiner}). In addition, there are also studies that employ the GTD formalism without adopting the most recent consistent version of GTD for quasi-homogeneous systems \cite{quevedo2023unified}. Instead, they rely on the original GTD formulation developed for homogeneous systems \cite{quevedo2007geometrothermodynamics}, as is the case in \cite{Hazarika, quevedo2008geometrothermodynamics, Gogoi}. Notably, \cite{larranaga2012geometric} employs GTD and does include the cosmological constant \( \Lambda \) as a thermodynamic quantity. However, their approach still differs from ours, as they do not utilize the quasi-homogeneous version of GTD and treat \( \Lambda \) itself, rather than \( l \), as the  thermodynamic variable. These distinctions can lead to significantly different thermodynamic geometries, phase structures, and microscopic behaviors.\\

As previously discussed, the divergences of \( \mathcal{R}^{II} \) coincides with the divergences of the heat capacity \( C_{lJ} \) and the compressibility \( \kappa_{S\Omega} \). However, \( \kappa_{S\Omega} \) diverges only for \( \Omega l \geq 1 \), but this regime may be thermodynamically excluded, as it violates the bound \( \Omega l < 1 \), which ensures a well-defined dual thermal state in the CFT and prevents superluminal rotation in the Einstein universe \cite{hawking1999rotation}. Nevertheless, in Figs.~ \ref{ff}(b) and \ref{ff}(d), we examine these singularities even in the excluded regime, as the GTD approach may provide valuable insights through its geometric correspondence with response functions. In Fig.~ \ref{ff}(b), we observe that in the region where \( T > 0 \), the scalar \( \mathcal{R}^{II} < 0 \), indicating that attractive thermodynamic interactions dominate for these Kerr-AdS black holes with small values of \( l \) and \( S \). This behavior, resembling the phase and microscopic structure of the Schwarzschild-AdS black hole, also shows purely attractive thermodynamic interactions \cite{Ladino:2024ned, ref14}. As seen in Fig.~ \ref{ff}(d), these divergences, corresponding to those of \( \kappa_{S\Omega} \), also occur for black holes with very high temperatures and when SBH and LBH are indistinguishable. Thus, these thermodynamic processes could be related to primordial or low-mass black holes \cite{Carr_2021, Greene_2012}, where this singularity may indicate a phase transition or signs of quantum effects in the microscopic behavior. \\

Our findings offer a novel perspective on the phase structure and microstructure of the Kerr-AdS black hole, presenting a clear contrast to Ruppeiner’s approach \cite{Hazarika, Banerjee, ref20, wei2021general, Zangeneh_2018, Dehyadegari, Sahay}. Notably, GTD inherently preserves Legendre invariance and accounts for the system’s quasi-homogeneity, thereby preventing unphysical singularities. Furthermore, it has been demonstrated that Ruppeiner’s theory encounters difficulties in effectively describing thermodynamic interactions for AdS black holes \cite{mirza2007ruppeiner}. For example, the Schwarzschild-AdS black hole, described in the equilibrium space \((S, l)\), exhibits a zero Ruppeiner's curvature, implying the absence of interactions, an outcome that contradicts black hole thermodynamics. This issue is typically addressed by interpreting the cosmological constant as pressure, and studying the system in the \((S, P)\) space, where interactions naturally emerge \cite{ref14}. Thus, Ruppeiner’s approach may face challenges in describing the Kerr-AdS black hole when \( l \) is treated as a thermodynamic variable, whereas GTD provides a more consistent framework.

\section{Shadow Thermodynamics}
In the following, we focus exclusively on the canonical ensemble. Initially, we review the behavior of the thermodynamic parameters of Kerr-AdS black holes at the critical point. In spherically symmetric and static spacetimes, it is possible to determine the radius of the critical photon sphere associated with a black hole whose temperature is critical and corresponds to an inflection point, as explained in \cite{Ladino:2024ned, WangRuppeiner}. In the case of Kerr-AdS black holes, however, we will see that there is no a single critical photon radius but rather a critical photon region.  To find the photon region in terms of the parameters $\{M,l,J\}$ instead of $\{m, l, a\}$, we rewrite the polynomial in Eq.~ \eqref{photon1} using Eq.~ \eqref{MandJ}. Then, considering the case where the curvature radius and black hole mass take the critical values \( l_c \equiv \bar{l} \sqrt{J} \) and \( M_c \equiv \bar{M} \sqrt{J} \), where \( \bar{l} \) and \( \bar{M} \) are the constant coefficients given in Eq.~ \eqref{critical1}, we obtain

\label{sec:shtd}

\begin{equation}
\begin{aligned}
 \frac{4\left(1-\bar{l}^2 \bar{M}^2\right)^2}{\bar{M}^6}\Big\{r^4+\bar{l}^2 J r\Big{[}r-2\left(1-\frac{1}{\bar{l}^2 \bar{M}^2}\right)^2 \bar{M} \sqrt{J}\Big{]}+\frac{J\left(\bar{l}^2 J+r^2\right)}{\bar{M}^2}&\Big\} \\
 -\Big\{\bar{l}^3 r\Big{[}r-3\left(1-\frac{1}{\bar{l}^2 \bar{M}^2}\right)^2 \bar{M} \sqrt{J}\Big{]}+\frac{\bar{l}\left(2 \bar{l}^2 J+r^2\right)}{\bar{M}^2}&\Big\}^2=0.
\end{aligned}
\label{photon2}
\end{equation}
The two largest real and positive roots of the polynomial above in $r$ determine the limiting values of the critical photon region $r_{sc}$. That is $r_{sc}^{-}\leq r_{sc}\leq r_{sc}^{+}$, which is approximately
\begin{equation}
4.4737 \sqrt{J} \lesssim r_{sc} \lesssim 6.7081\sqrt{J}.
\label{photon3}
\end{equation}
Hence, the critical photon region, as expected, depends only on $J$, similar to the other critical values of Eq.~ \eqref{critical1}. Next, we proceed to calculate the critical quantities for the impact parameters and the celestial coordinates that describe the projection of the shadow boundary.
\subsection{Thermodynamics through shadow size and profiles}
First, we examine the critical thermodynamic quantities through the shadow, and then we  conduct an analysis of the thermodynamic properties derived from asymmetric shadow profiles and the behavior of the shadow radius. Thus, to rewrite the impact parameters and the celestial coordinates of Eq.~ \eqref{impactfactors}, \eqref{celestial1} and \eqref{celestial2} in terms of the parameters $\{M,l,J\}$ instead of $\{m, l, a\}$, we use Eq.~ \eqref{MandJ}. Subsequently, by considering the critical photon region as \( r_{sc} \equiv \bar{r} \sqrt{J} \), where \( \bar{r} \) is the constant coefficient given in the interval of Eq.~ \eqref{critical1}, and substituting the critical values \( l_c \equiv \bar{l} \sqrt{J} \) and \( M_c \equiv \bar{M} \sqrt{J} \), the critical impact parameters take the form (see also \cite{tsukamoto2014constraining,tsukamoto2018black})
\begin{align}
    \xi_c&=  \frac{1-\bar{l}^2 \bar{M}^2\sqrt{J}}{\bar{M} \mathcal{C}}\Big\{\bar{M}\left[3 \bar{M} \bar{r}^2+\bar{l}^2\left(\bar{M}+\bar{r}-3 \bar{M}^3 \bar{r}^2+\bar{M}^2 \bar{r}^3\right)\right]-1\Big\}   ,\label{criticalimpact1}\\
    \eta_c&=  \frac{4\left(1-\bar{l}^2 \bar{M}^2\right)^2 \bar{r}^2 J}{\bar{M} \mathcal{C}^2}\Big\{\bar{l}^4 \bar{M}\left[1-\bar{M}^2(2 \bar{M}-\bar{r}) \bar{r}\right]+\bar{l}^2 \bar{M} \bar{r}\left[\bar{r}+\bar{M}\left(4+\bar{M} \bar{r}^3\right)\right]-2 \bar{r}\Big\}. \label{criticalimpact2}
\end{align}
 Here, \( \mathcal{C} \) is a constant defined in Appendix \ref{apenA}, 
 Eq.~ \eqref{Cequation}. Using the above equations, the explicit expressions for the critical celestial coordinates \( \rho_c \) and \( \sigma_c \) of the shadow are given in Eqs.~ \eqref{critical2} and \eqref{critical3} of Appendix \ref{apenA}. Substituting these critical values into the cartesian coordinates from Eqs.~ \eqref{cartecian} and applying Eqs.~ \eqref{radiusshadow} and \eqref{photon3}, we numerically obtain the critical shadow radius \( R_{shc} \). This critical value \( R_{shc} \) depends only on the angular momentum \( J \) and the observer's radial position \( r_0 \). Moreover, this result is consistent with our previous findings on the critical shadow radius of the RN-AdS black hole \cite{Ladino:2024ned}, which similarly depends on the electric charge \( Q \) and the observer's radial position \( r_0 \).  
\begin{figure}[H]
  \centering
  \begin{subfigure}{0.42\textwidth}
  \caption*{(a) $\hspace{6.6cm}$ $\vspace{-0.2cm}$ }
    \includegraphics[width=\linewidth]{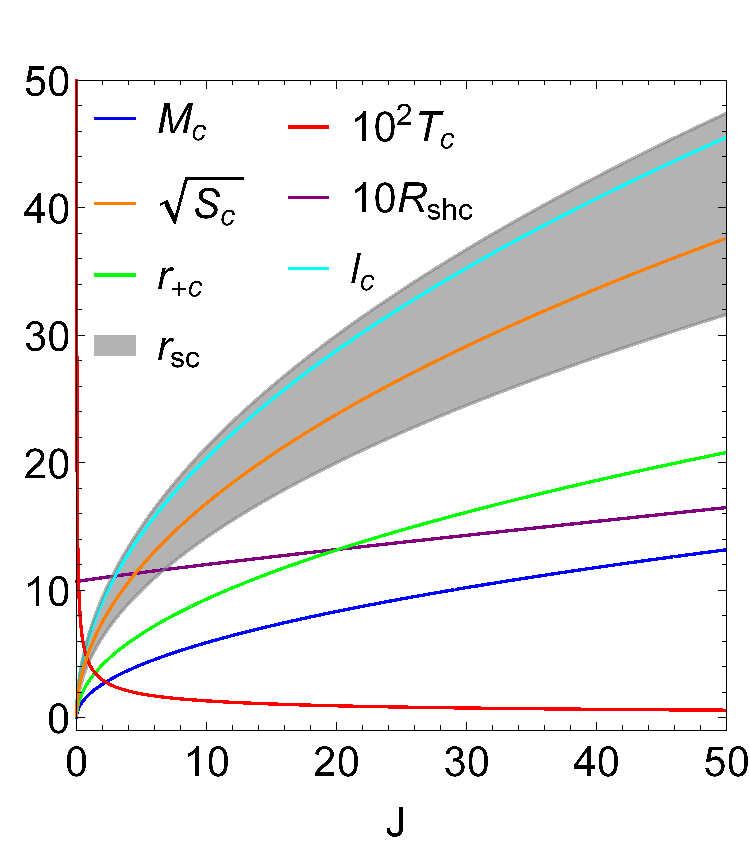}
  \end{subfigure}
  \hfill
  \begin{subfigure}{0.47\textwidth}
  \caption*{(b) $\hspace{6.6cm}$ $\vspace{0.5cm}$}
    \includegraphics[width=\linewidth]{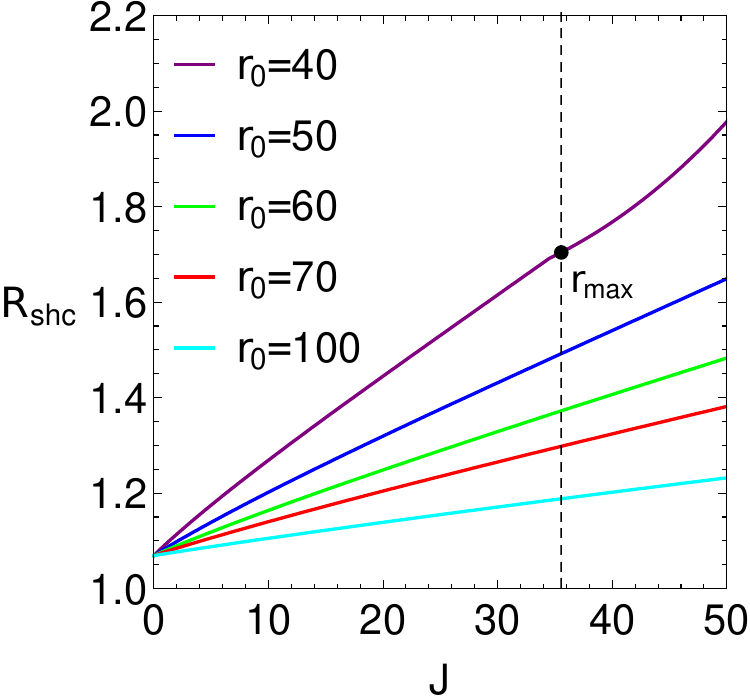}
  \end{subfigure}
  \caption{(a) Behavior of critical quantities in terms of the angular momentum $J$ ($R_{shc}$ is calculated for $r_0=100$). (b) Critical shadow radius $R_{shc}$ as a function of the angular momentum $J$ for various values of the observer's location $r_0$. $r_{max}$ indicates the possible maximum value
for $R_{shc}$.}
  \label{fig:critical}
\end{figure}
In Fig.~ \ref{fig:critical}, we show the behavior of the critical quantities from Eq.~ \eqref{critical1}, along with the critical photon region (shaded in gray) and the critical shadow radius $R_{shc}$, all as functions of the angular momentum $J$. All critical quantities exhibit a similar increasing behavior, except for the critical temperature $T_{c}$, which diverges at the origin and decreases as $J$ increases. We can observe that, for a fixed $J$, the critical shadow radius $R_{shc}$ decreases as the observer's distance $r_0$ increases. Conversely, for a fixed $r_0$, $R_{shc}$ increases as $J$ grows. However, as shown in \cite{Ladino:2024ned}, $R_{shc}$ does not grow indefinitely; instead, it reaches a maximum value, denoted as $r_{max} \approx 1.7043$, which corresponds to the case  when the photon region reaches $r_0$. In fact, by considering  $J \rightarrow 0$ as the lower bound and $r_{sc} \rightarrow r_0$ as the upper bound, we can infer some approximate constraints, such as
\begin{equation}
 1.0692  \lesssim R_{shc}\lesssim1.7043,\quad \quad\text{for}\quad \quad\ 0\lesssim J\lesssim0.0222r_0^2 .
 \label{restriccion1}
\end{equation}
Consequently, the shadow radius reaches a critical value $R_{shc}$ when the black hole undergoes a second-order phase transition, satisfying the limits established in Eq.~ \eqref{restriccion1}. This critical value is influenced by the observer's distance, $r_0$. As a result, a Kerr-AdS black hole with a shadow radius in the range $1.0692 \lesssim R_{sh} \lesssim 1.7043$ could represent a second-order phase transition. Additionally, the value $R_{shc} \approx 1.0692$ corresponds to the shadow radius of a Schwarzschild-AdS black hole that could undergo a second-order phase transition. However, it is important to note that, in this case, the values of $R_{sh}$ are dimensionless. This is due to the stereographic projection of Eq.~ \eqref{cartecian} used to obtain the cartesian coordinates $x$ and $y$, which are also dimensionless. Therefore, caution is needed when comparing these results with radii obtained using other approaches or in different spacetimes. An additional coordinate transformation can be performed to convert to Bardeen's dimensional variables \cite{Perlick_2022}, $(\alpha, \beta)$, using $\alpha = r_0 x - a \sin(\theta_0)$ and $\beta=r_0 y$. However, this transformation is only valid when $r_0 \gg m$, as Bardeen's method is applicable only to observers located far from the black hole. In our case, since our spacetime is asymptotically AdS, this approximation is not suitable, as it requires the observer to be at a finite distance less than the cosmological horizon, rather than at infinity. 
\begin{figure}[H]
  \centering
  \begin{subfigure}{0.44\textwidth}
  \caption*{(a) \hspace{0.3cm}  $J=1$, $l=5$, $r_0=100$, $\theta_0=\pi/2$
  $\vspace{0.5cm} $}
    \includegraphics[width=\linewidth]{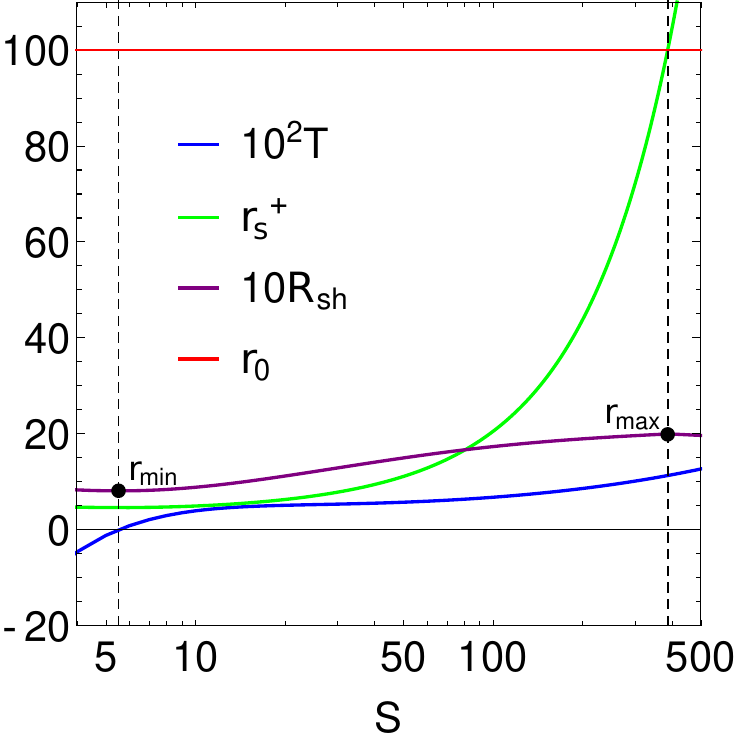}
  \end{subfigure}
  \hfill
  \begin{subfigure}{0.47\textwidth}
  \caption*{(b) \hspace{0.3cm}  $J=1$, $l=5$, $r_0=100$, $\theta_0=\pi/2$}
    \includegraphics[width=\linewidth]{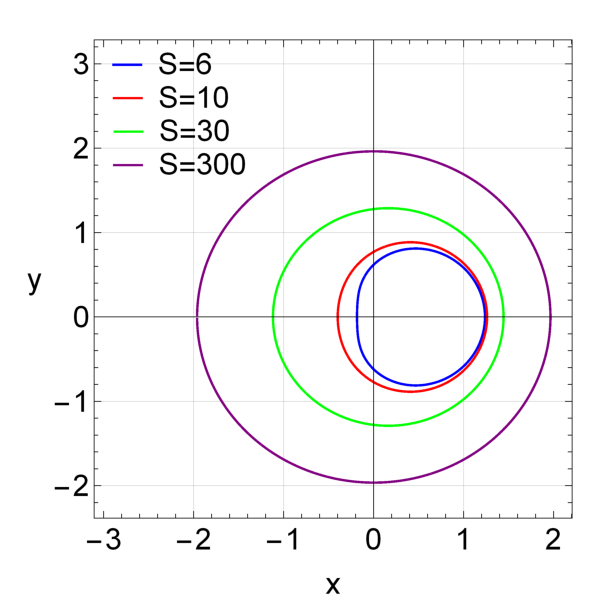}
  \end{subfigure}
  \caption{(a) Temperature $T$, radius of the upper limit of the photon region $r_s^+$, shadow radius $R_{sh}$, and observer location $r_0$ as functions of the entropy $S$. $r_{min}$ and $r_{max}$ indicate the possible limiting values for $R_{sh}$. (b) Shadow cast behavior for different values of the entropy $S$.}
  \label{fig:figura2_completa}
\end{figure}
In Fig.~ \ref{fig:figura2_completa}, the left panel presents the temperature $T$, the upper limit of the photon region $r_s^+$, the shadow radius $R_{sh}$, and the observer's location $r_0$ as functions of the entropy $S$. One of the main ideas of the shadow thermodynamics formalism is to represent thermodynamic properties as functions of the shadow radius $R_{sh}$ or to project them onto the black hole's shadow cast. To achieve this correctly, we must carefully ensure a faithful physical representation of the shadow. Therefore, for the Kerr-AdS black hole, we can identify two local extremum points, $r_{min}$ and $r_{max}$, which divide the shadow curve into three regions. Only the segment $r_{min} < R_{sh} < r_{max}$ is physically relevant, as it reflects the phase transition of the Kerr-AdS black hole. The other two segments are non-physical: for $R_{sh} < r_{min}$, the temperature is negative ($T < 0$), while for $R_{sh} > r_{max}$, the case is no longer applicable, since the observer is always assumed to be located outside the photon region (\( r_{s}<r_{0} \)) \cite{Ladino:2024ned, WangRuppeiner}. \\

While several recent studies have explored the relationship between thermodynamic properties and black hole shadows in spherical cases, focusing on projected profiles of the critical shadow curve to reveal the behavior of quantities such as temperature and heat capacity (as it was done in \cite{Luo, GuoLi, Belhaj, WangRuppeiner, Du, Zheng, He, ref21, Ladino:2024ned}), this work extends the analysis to rotating black holes. By examining thermodynamic properties through asymmetric shadow profiles, we provide new insights into the connection between rotating black hole thermodynamics and its measurable characteristics.
$\vspace{-0.2cm}$
\begin{figure}[H]
  \centering
  \begin{subfigure}{0.49\textwidth}
    \caption*{ (a) $\vspace{-0.2cm}$$\hspace{7.7cm}$ }
    \includegraphics[width=\linewidth]{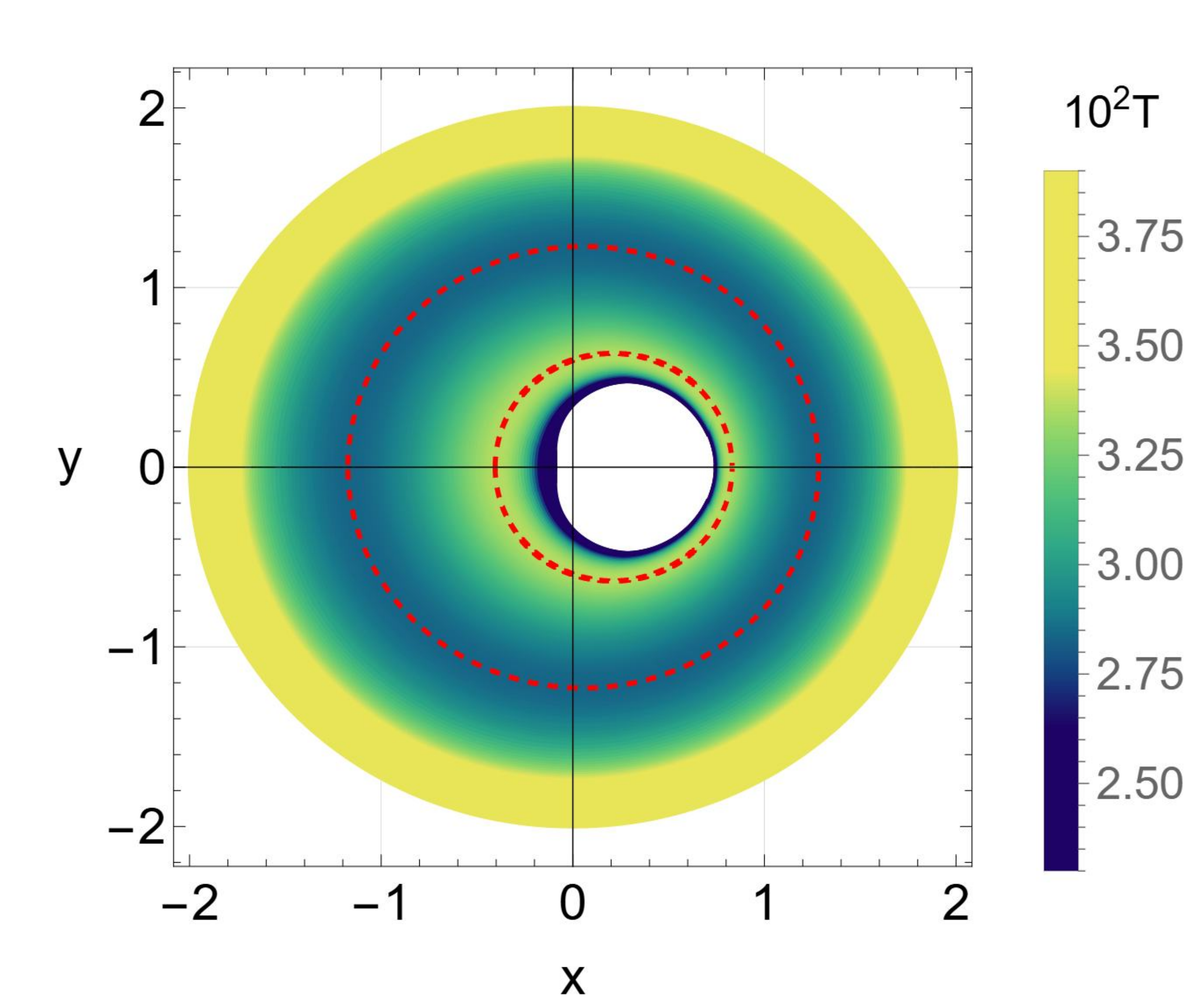}
  \end{subfigure}
  \hfill
  \begin{subfigure}{0.485\textwidth}
    \caption*{ (b) $\vspace{-0.2cm}$$\hspace{7.7cm}$ }
    \includegraphics[width=\linewidth]{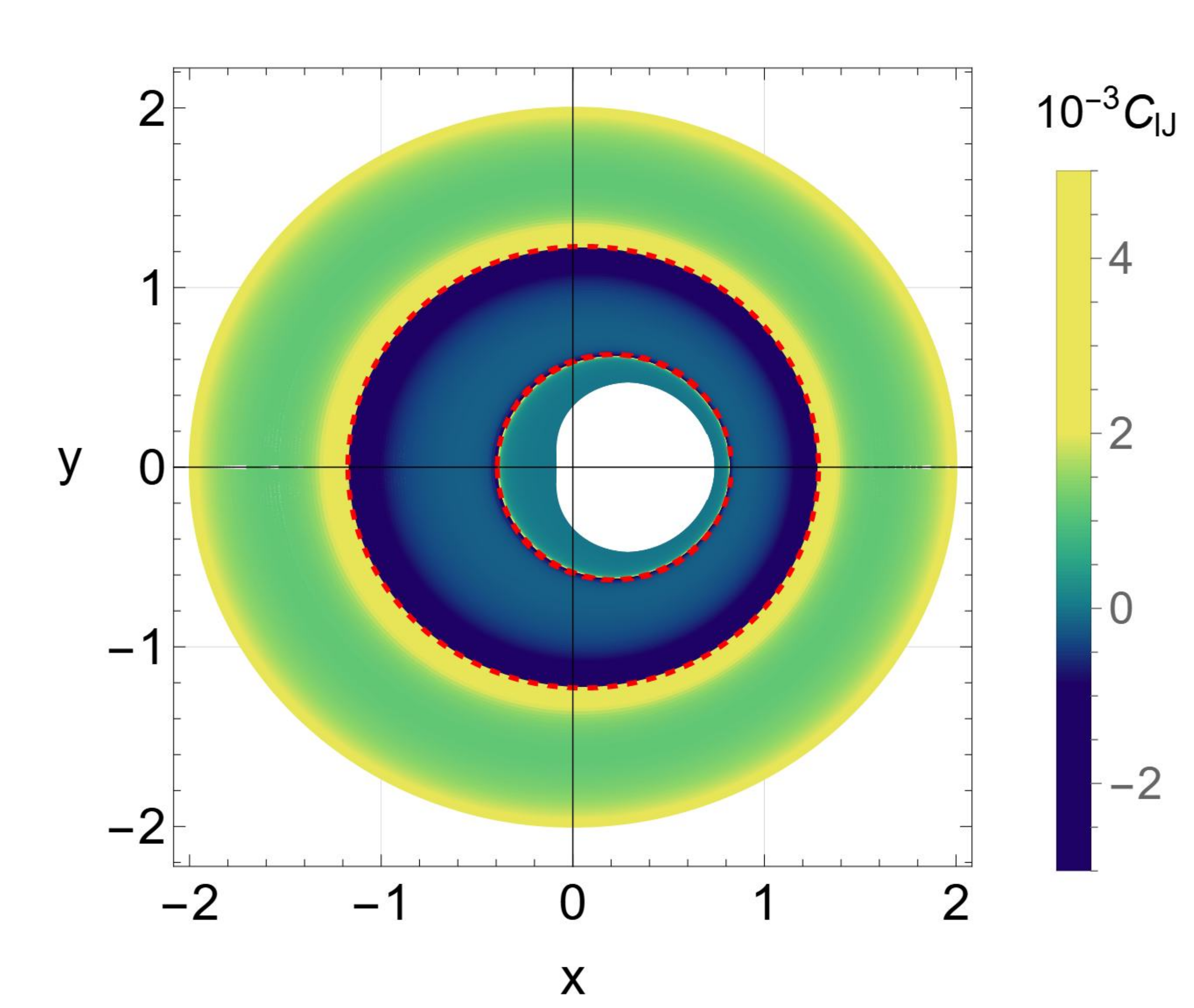}
  \end{subfigure}
    \centering
       $\vspace{-1.7cm}$
  \begin{subfigure}{0.39\textwidth}
    $\hspace{0.05cm}$
   $\vspace{-0.01cm}$
    \includegraphics[width=\linewidth]{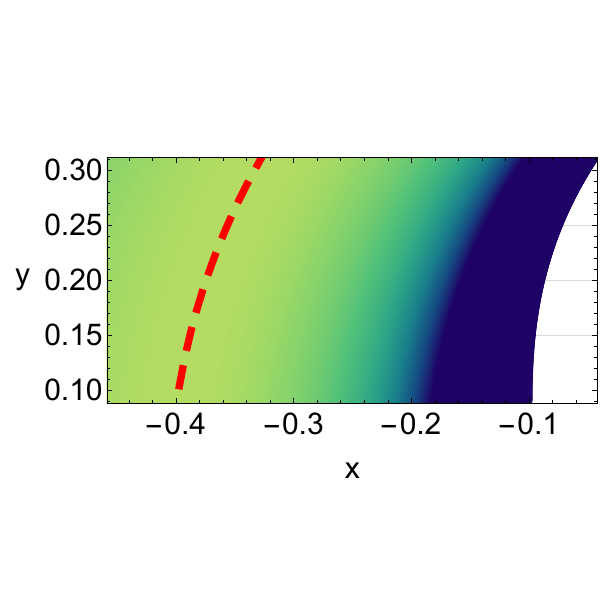}
  \end{subfigure}
  $\vspace{-0.4cm}$
  \hfill
  \begin{subfigure}{0.385\textwidth}
  $\hspace{-1.95cm}$
   $\vspace{-0.01cm}$
    \includegraphics[width=\linewidth]{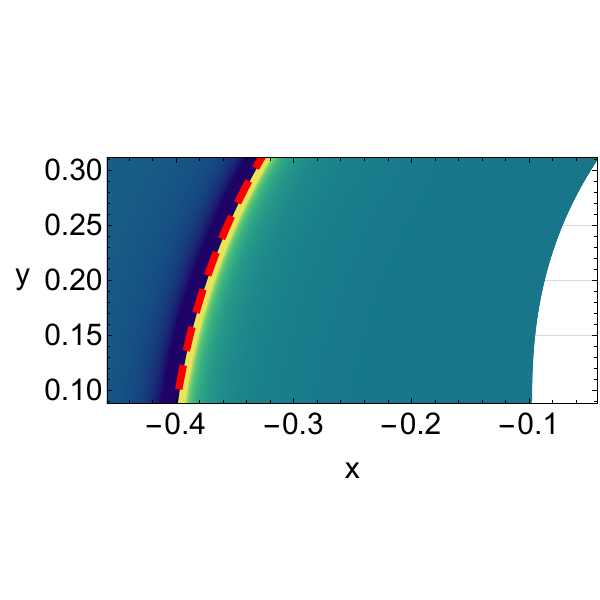}
  \end{subfigure}
  $\vspace{-0.8cm}$
  \caption{(a) Temperature $T$ and (b) heat capacity \( C_{lJ} \) across the shadow’s cast profile of Kerr-AdS black holes (for $l=1.5l_c$, $J=1$, $r_0=100$ and $\theta_0=\pi/2$). The red dotted lines indicate the critical points of \( T \) and the singularities of \( C_{lJ} \). The bottom panels provide a zoomed-in view of the corresponding shadow case from the upper panel, highlighting the region of the first critical point and singularity where both occur.}
  \label{fig:figura0_completa3}
\end{figure}
The right panel of Fig.~\ref{fig:figura2_completa} shows the shadow for different entropy values $S$, obtained by expressing the celestial coordinates from Eqs.~~\eqref{celestial1} and \eqref{celestial2} in terms of $\{S, l, J, r_0, \theta_0\}$, instead of \( \{m, l, a, r_0, \theta_0\} \), via Eqs.~ \eqref{MandJ} and \eqref{fundame1}.  Thus, for fixed \( J \) and \( l \), we can find a valid region of \( S \) values for which the shadow radius satisfies \( r_{min} < R_{sh} < r_{max} \).  Based on this, we create asymmetric shadow profiles of the thermodynamic quantities by overlaying critical curves. Fig.~ \ref{fig:figura0_completa3} displays a color intensity map over the shadow, representing the variation of the temperature \( T \) and heat capacity \( C_{lJ} \) across the Kerr-AdS black hole’s shadow profile. This visualization reveals the black hole’s phase structure by highlighting the critical points of \( T \) and the singularities of \( C_{lJ} \) along overlapping critical curves, confirming the connection between shadow size and thermodynamic phase transitions.
\begin{figure}[H]
  \centering
  \begin{subfigure}{0.31\textwidth}
  \caption*{ $ J=0.3$}
    \includegraphics[width=\linewidth]{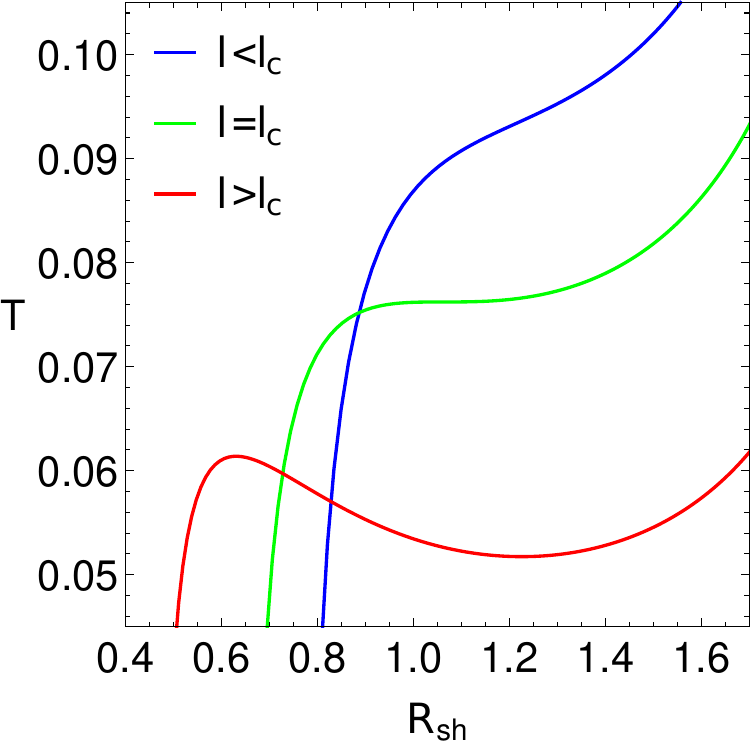}
  \end{subfigure}
  \hfill
  \begin{subfigure}{0.31\textwidth}
      \caption*{ $ J=0.6$}
    \includegraphics[width=\linewidth]{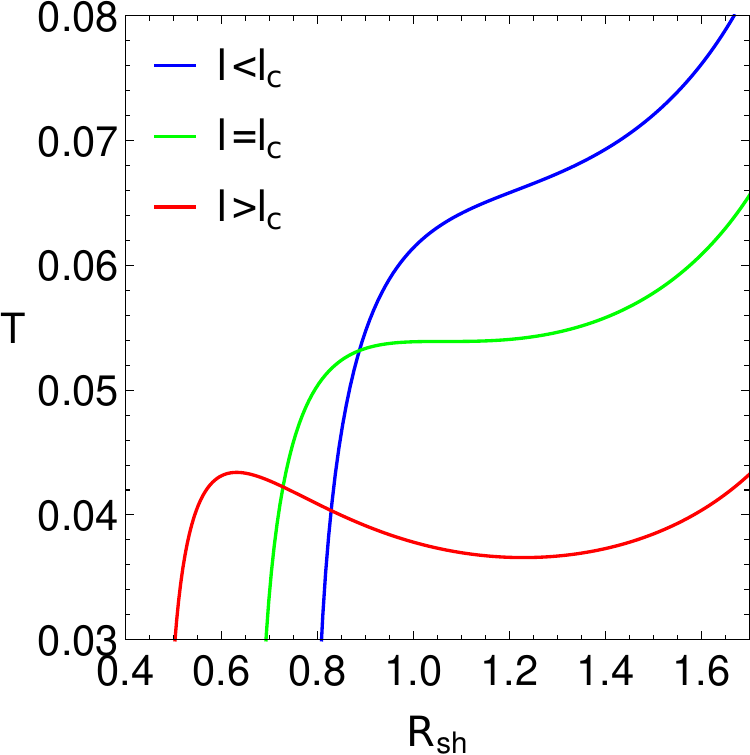}
  \end{subfigure}
  \hfill
  \begin{subfigure}{0.31\textwidth}
      \caption*{ $ J= 1$}
    \includegraphics[width=\linewidth]{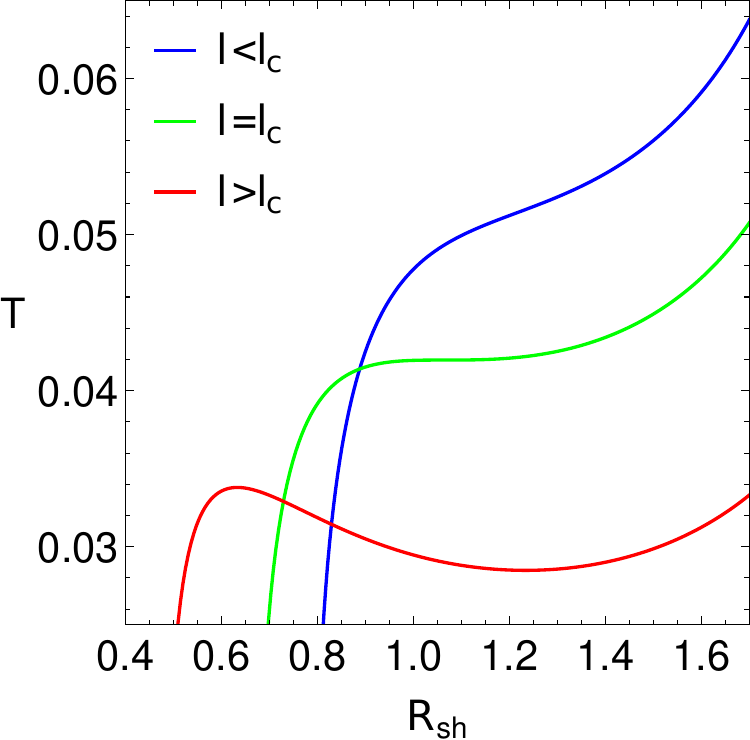}
  \end{subfigure}
  \vspace{0.5cm}
    \centering
  \begin{subfigure}{0.325\textwidth}
  \caption*{ $ J=0.3$}
    \includegraphics[width=\linewidth]{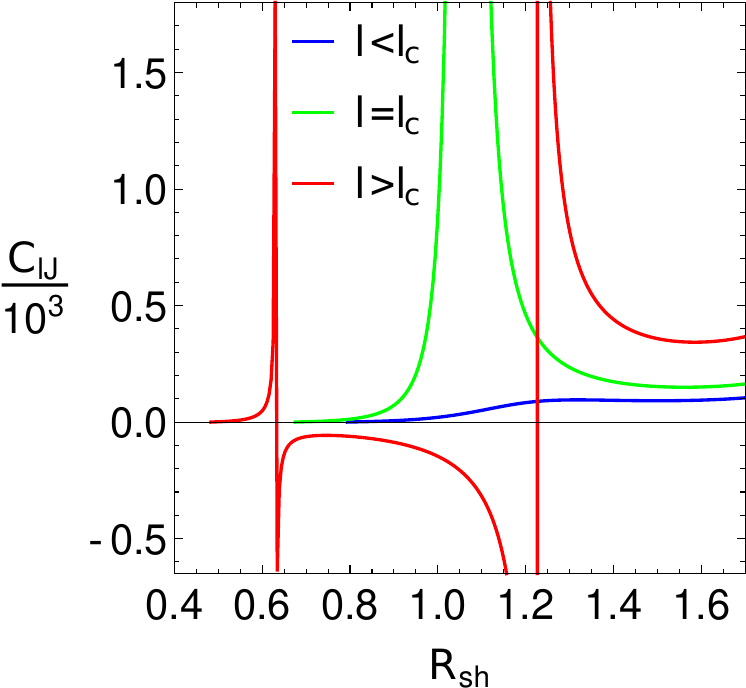}
  \end{subfigure}
  \hfill
  \begin{subfigure}{0.325\textwidth}
      \caption*{ $ J=0.6$}
    \includegraphics[width=\linewidth]{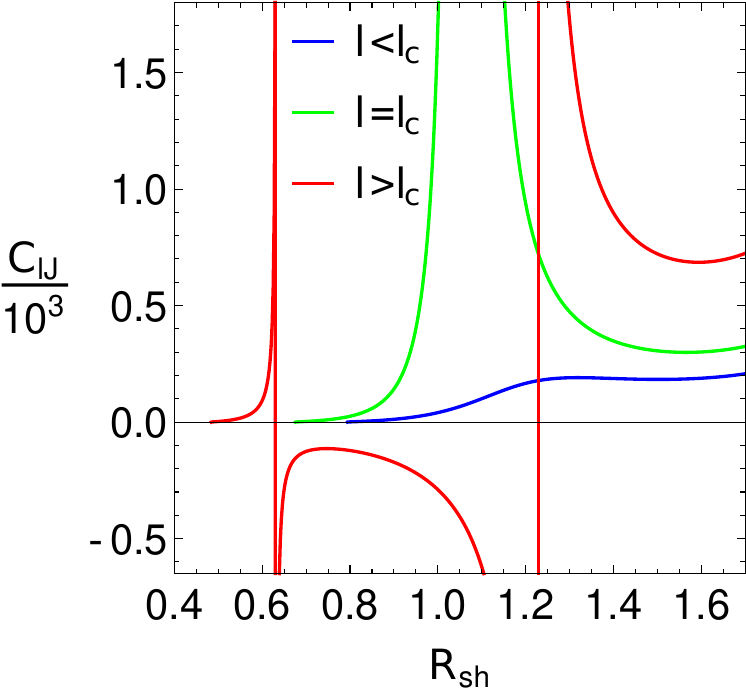}
  \end{subfigure}
  \hfill
  \begin{subfigure}{0.325\textwidth}
      \caption*{ $ J= 1$}
    \includegraphics[width=\linewidth]{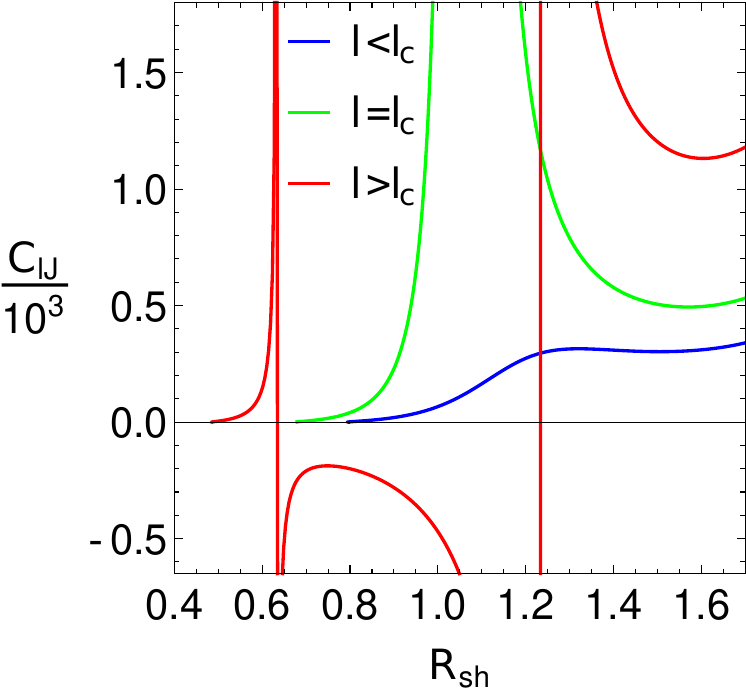}
  \end{subfigure}
  \caption{Temperature $T$ (top) and heat capacity $C_{lJ}$ (bottom) as functions of the shadow radius $R_{sh}$ for various values of $J$ and $l$. Here, $r_0=100$ and $\theta_0=\pi/2$.}
  \label{fig:figura1_completa}
\end{figure}
In Fig.~ \ref{fig:figura1_completa}, we illustrate the behavior of the temperature \( T \) and heat capacity \( C_{lJ} \) as functions of the shadow radius \( R_{sh} \) for different values of \( J \) and \( l \). There, by comparing with the results shown in Fig.~ \ref{fig:figura11_completa}, we can corroborate that the shadow radius \( R_{sh} \) correctly reflects the phase transition configuration of the Kerr-AdS black hole. As discussed in previous sections regarding Fig.~ \ref{fig:figura11_completa} in the canonical ensemble, for \( l > l_c \), an intermediate unstable branch emerges, indicating the coexistence of small black holes in a liquid phase and large black holes in a gaseous phase, undergoing a first-order phase transition. At \( l = l_c \), a critical inflection point in \( T \) and a single singularity in \( C_{lJ} \) lead to a second-order phase transition. For \( l < l_c \), the black hole is in a stable supercritical phase, where the small and large phases become indistinguishable \cite{wang2022ruppeiner}. Results that the shadow radius \( R_{sh} \) faithfully shows. \\

Unlike the usual analysis based on entropy \( S \) or the horizon radius \( r_+ \), the phase transition configuration in terms of \( R_{sh} \) is quite similar across different values of the angular momentum \( J \). For example, the divergences in \( C_{lJ} \) are nearly indistinguishable for different values of \( J \). This suggests that once the angular momentum $J$ of a Kerr-AdS black hole is identified from its shadow, it becomes challenging to obtain precise information about its thermodynamic properties, as the spin may not be a determining factor of its thermodynamic state (as long as it is not identically zero). Conversely, once the shadow radius of the black hole is determined, it would be difficult to associate a specific thermodynamic phase with a particular angular momentum. On the other hand, changes in the curvature radius \( l \) suggest that this could be a determining parameter of the thermodynamic state of the Kerr-AdS black hole as seen from its shadow. This makes sense when we recall that the spin of a black hole has little effect on the shadow radius, even for high inclination angles, where the effect remains \(\lesssim 12\) \cite{Vagnozzi_2023}.

\subsection{Geometrothermodynamics of black hole shadows}
\label{sec:shgtd}
The singularity condition \( II \) from Eq.~ \eqref{qa}, associated with the metric \( g^{II} \) in Eq.~ \eqref{metric2} within the GTD formalism, coincides with the divergences of the heat capacity \( C_{lJ} \) from Eq.~ \eqref{temp4}. Building on this, we analyze the thermodynamic phase structure of the Kerr-AdS black hole through its shadow using the GTD metric \( g^{II} \). This approach provides a novel perspective on the connection between black hole geometric thermodynamics and its observational features.\\

\begin{figure}[H]
  \centering
  \begin{subfigure}{0.5\textwidth}
    \includegraphics[width=\linewidth]{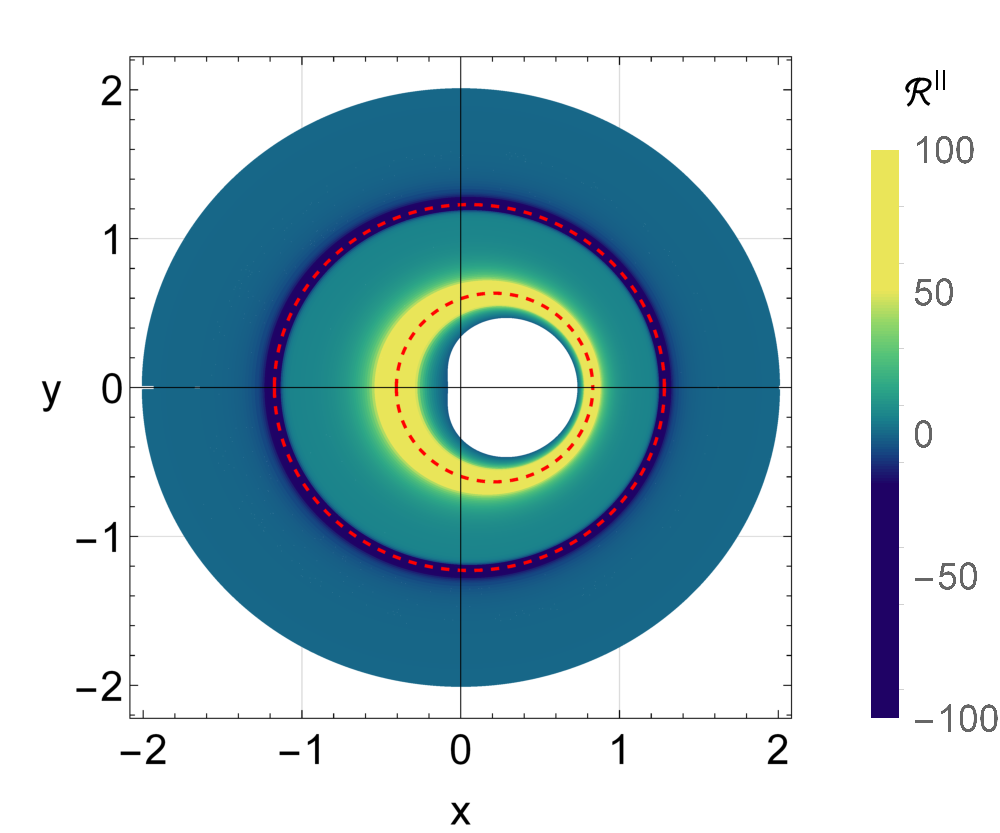}
    $\vspace{-1.7cm}$
    \caption*{ (a) $\hspace{8.7cm}$ }
  \end{subfigure}
  \hfill
  \begin{subfigure}{0.38\textwidth}
  $\hspace{-1.2cm}$$\vspace{0.6cm}$   \includegraphics[width=\linewidth]{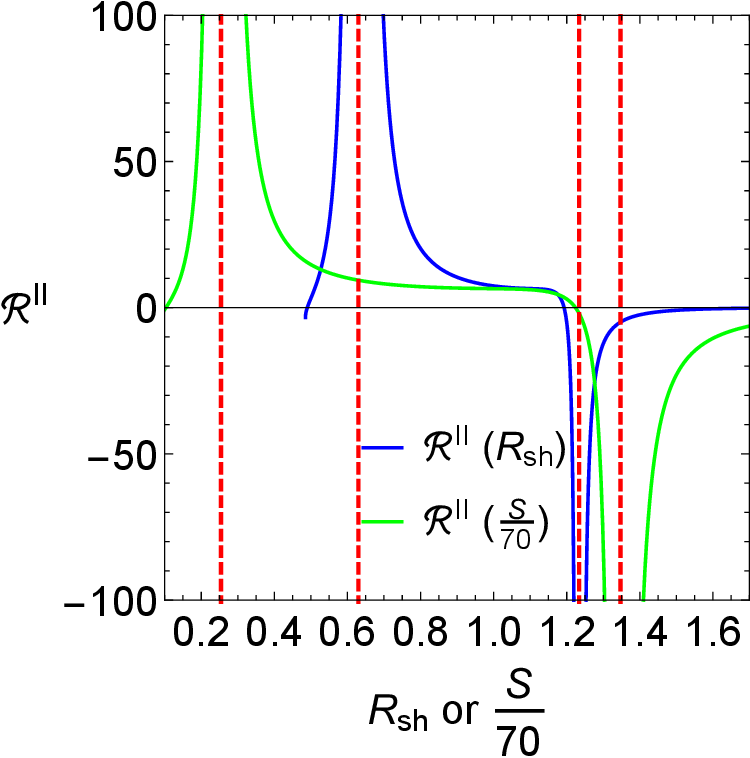}
        $\vspace{-1.7cm}$
    \caption*{ (b) $\hspace{9.7cm}$ }
  \end{subfigure}
   \begin{subfigure}{0.5\textwidth}
    \includegraphics[width=\linewidth]{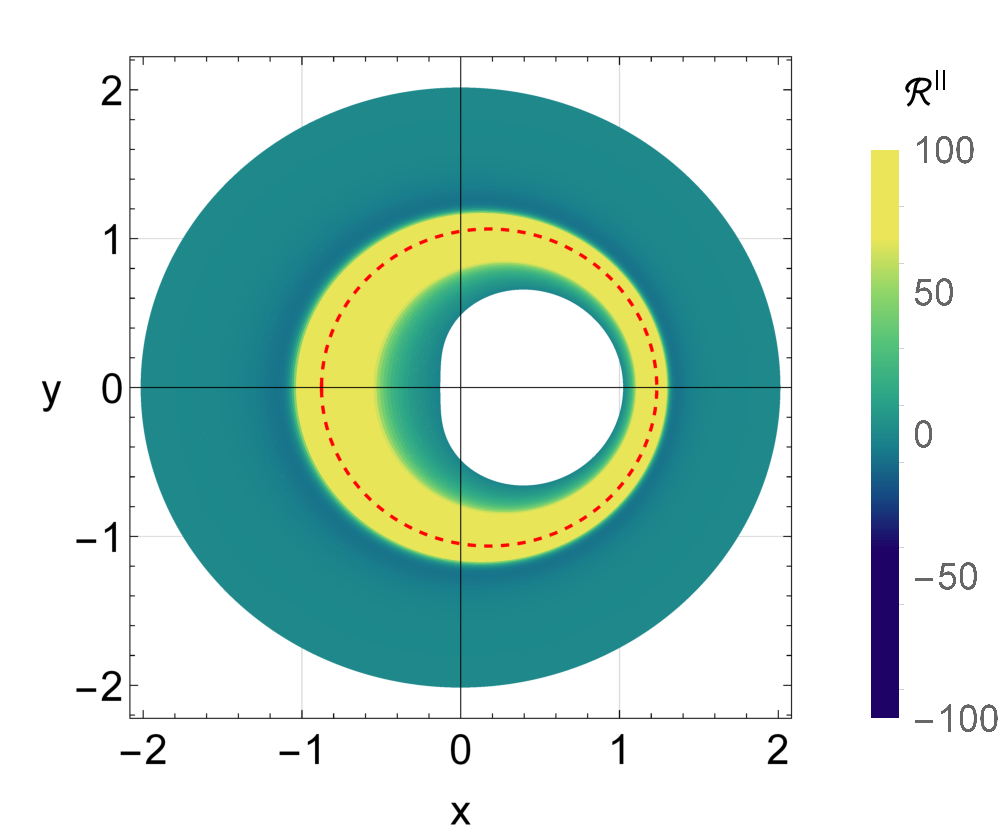}
        $\vspace{-1.7cm}$
    \caption*{ (c) $\hspace{8.7cm}$ }
  \end{subfigure}
  \hfill
  \begin{subfigure}{0.38\textwidth}
  $\hspace{-1.2cm}$$\vspace{0.6cm}$    \includegraphics[width=\linewidth]{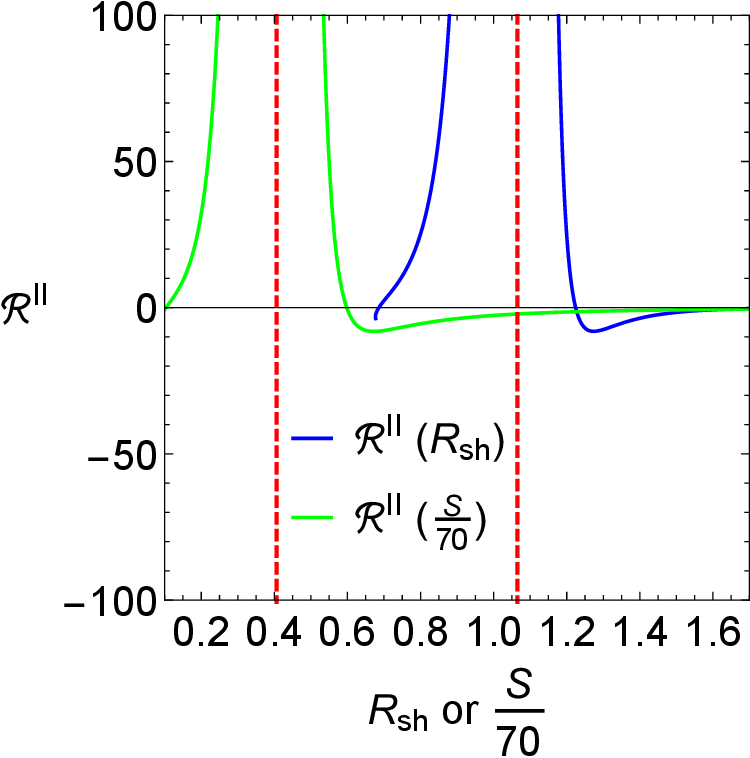}
        $\vspace{-1.7cm}$
     \caption*{ (d) $\hspace{9.7cm}$ }
  \end{subfigure}
    \begin{subfigure}{0.5\textwidth}
    \includegraphics[width=\linewidth]{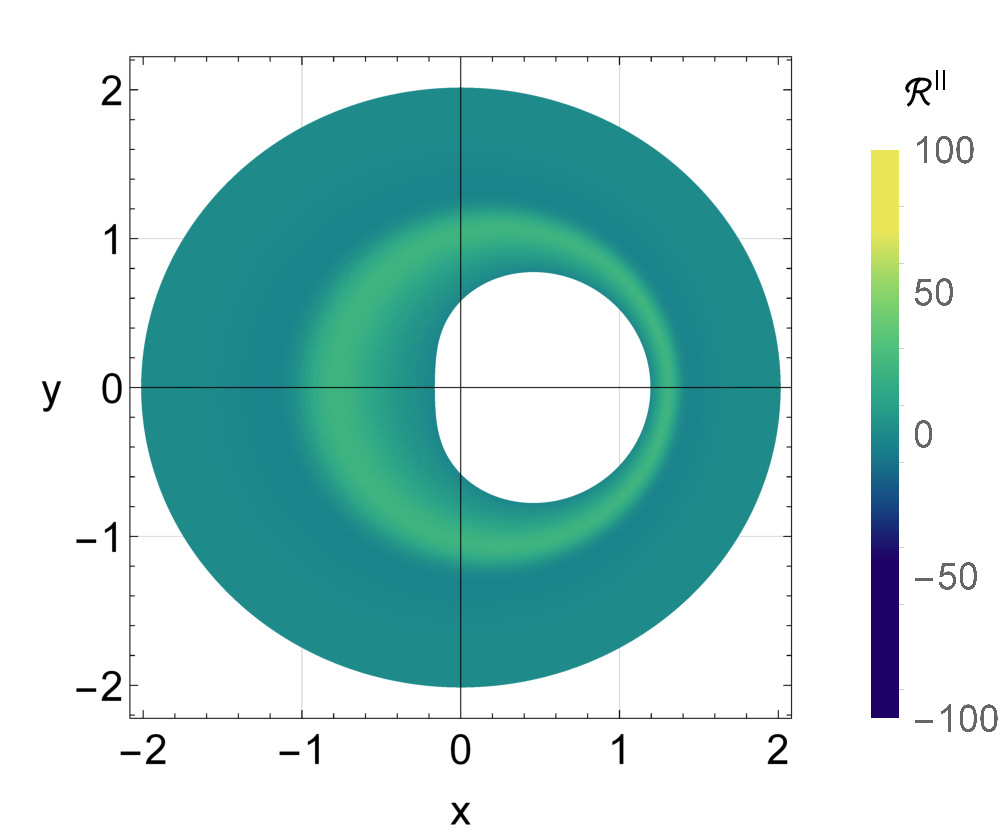}
        $\vspace{-1.7cm}$
    \caption*{ (e) $\hspace{8.7cm}$ }
  \end{subfigure}
  \hfill
  \begin{subfigure}{0.38\textwidth}
  $\hspace{-1.2cm}$$\vspace{0.6cm}$    \includegraphics[width=\linewidth]{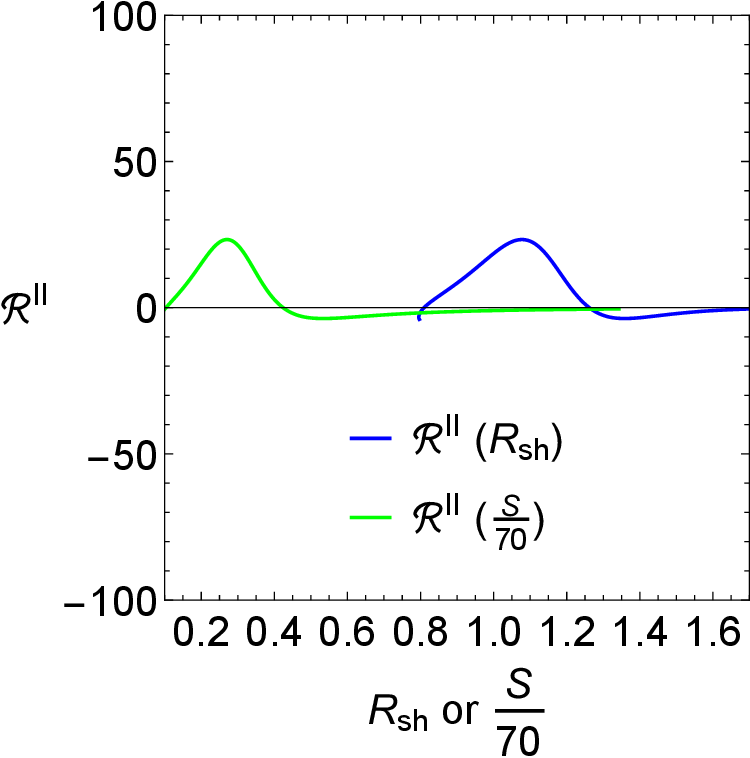}
        $\vspace{-1.7cm}$
     \caption*{ (f) $\hspace{9.7cm}$ }
  \end{subfigure}
  $\vspace{0.4cm}$
  \caption{ Variation of the scalar $\mathcal{R}^{II}$ across the shadow’s cast profile and $\mathcal{R}^{II}$ as functions of the shadow radius $R_{sh}$ and entropy $S$ for Kerr-AdS black holes. The red dotted lines indicate the divergences of $\mathcal{R}^{II}$. (a) and (b) for $l=1.5l_c$, (c) and (d) for $l=l_c$ and (e) and (f) for $l=0.8l_c$. Here, $J=1$, $\beta_M=1$, $r_0=100$ and $\theta_0=\pi/2$.}
  \label{fig:scalarprofile1}
\end{figure}

To analyze the behavior of the scalar \( \mathcal{R}^{II} \) associated with the GTD metric \( g^{II} \), we generate the shadow cast profiles of the Kerr-AdS black hole using the same methodology as in Fig.~ \ref{fig:figura0_completa3}, fixing \( J \) and \( l \) and considering a valid range of \( S \) values for which the shadow radius satisfies \( r_{\text{min}} < R_{\text{sh}} < r_{\text{max}} \). Accordingly, the left panels of Fig.\ref{fig:scalarprofile1} illustrate the variation of the Ricci scalar \( \mathcal{R}^{II} \) across the Kerr-AdS black hole’s shadow profile, highlighting the phase transition configurations along overlapping critical curves. Complementarily, the right panels of Fig.\ref{fig:scalarprofile1} show \( \mathcal{R}^{II} \) as a function of the shadow radius \( R_{\text{sh}} \) and entropy \( S \), further confirming the connection between shadow properties and thermodynamic phase structure. This thermodynamic configuration exhibits first-order and second-order phase transitions depending on the curvature radius \( l \). For \( l > l_c \) (Figs.\ref{fig:scalarprofile1}(a) and \ref{fig:scalarprofile1}(b)), two singularities emerge, indicating a first-order phase transition. For \( l = l_c \) (Figs. \ref{fig:scalarprofile1}(c) and \ref{fig:scalarprofile1}(d)), a single singularity appears, corresponding to a second-order phase transition. Finally, for \( l < l_c \) (Figs. \ref{fig:scalarprofile1}(e) and \ref{fig:scalarprofile1}(f)), no phase transitions occur. This behavior aligns with changes in the shadow radius. These figures demonstrate that the divergences of $\mathcal{R}^{II}$ are accurately captured by both $R_{sh}$ and $S$, corresponding to those of the heat capacity \( C_{lJ} \). \\

In the left panels of Fig.~\ref{fig:scalarprofile1}, the projected profiles of the critical shadow curve show that in the near-extremal case (\( T \approx 0 \)), the shadow takes on a distinct D-shape, especially for small radii \( R_{\text{sh}} \), a well-known feature of rotating black holes \cite{grenzebach2016shadow}. Each shadow profile also reveals the phase transitions of the black hole at fixed \( J \) and \( l \). Since \( R_{\text{sh}} \) increases with entropy \( S \) and the horizon radius \( r_+ \), each value of the scalar \( \mathcal{R}^{II} \) and its singularities corresponds to a specific \( R_{\text{sh}} \). Additionally, the phase transitions of Kerr-AdS black holes, associated with \( C_{lJ} \), occur for relatively medium-sized black holes, as the singularities are distant from the cases when \( T \approx 0 \) and \( r_s \approx r_0 \), which correspond to the smallest and largest \( R_{\text{sh}} \), respectively. This behavior is similar to that observed in the shadow profiles of RN-AdS black holes in \cite{Ladino:2024ned}, where a comparable phase structure was found, exhibiting an analogy with the gas-liquid phase transition in van der Waals systems \cite{callen1998thermodynamics}. Thus, the size of the shadow of the Kerr-AdS black hole faithfully reflects its thermodynamic phase structure under the GTD approach. Moreover, since the behavior of \( \mathcal{R}^{II} \) as a function of \( S \) correctly aligns with what is shown by \( R_{\text{sh}} \), we can conclude that the shadow size of Kerr-AdS black holes also reflects their black hole microstructure properties. Findings that had previously been established within the framework of Ruppeiner's thermodynamic geometry \cite{WangRuppeiner, ref20, ref21}. 
Compared to Ruppeiner's approach, GTD offers an innovative perspective for describing the thermodynamic properties that can be inferred from the shadows of Kerr-AdS black holes. As discussed in the previous section, the results obtained from GTD reveal a phase structure and black hole microstructure that differ from those described by Ruppeiner \cite{Hazarika, Banerjee, ref20, wei2021general, Zangeneh_2018, Dehyadegari, Sahay}. Moreover, GTD provides several key advantages over Ruppeiner's geometric formalism, including Legendre invariance, and the ability to treat the curvature radius \( l \) as a thermodynamic variable without resorting to any amount of pressure \cite{Ladino:2024ned}. These features make GTD a useful framework for analyzing the thermodynamic aspects of black hole shadows.\\

For \( a < l \), \( T > 0 \), and \( r_s > r_+ \), the Ricci scalar \( \mathcal{R}^{II} \) exhibits singularities at small values of \( l \) and \( S \), which are related to the compressibility singularity \( \kappa_{S\Omega} \) from Eq.~ \eqref{kappasomega}. These singularities may be unphysical, as they occur in regions where \( \Omega l > 1 \), as previously discussed in Fig.~ \ref{jcritical}(c). However, as shown in Figs.~ \ref{ff}(b) and \ref{ff}(d), they could provide valuable insights into the geometric correspondence of GTD with thermodynamic response functions or other possible theoretical interpretations. Nevertheless, the method used in this work to analyze black hole shadows is not suitable for studying these singularities, which appear at higher temperatures than those associated with \( \mathcal{R}^{II} \) and \( C_{lJ} \). Near this singularity of \( \kappa_{S\Omega} \), the celestial coordinates \( \rho \) and \( \sigma \) in Eqs.~ \eqref{celestial1} and \eqref{celestial2} often become complex. Even when \( \rho \) and \( \sigma \) remain real, the condition \( r_s > r_0 \) prevents the full projection of the critical shadow curve. It would be valuable to investigate whether this singularity appears in the shadows of rotating AdS black holes in alternative gravity theories, or how analyzing the full spectrum of singularities in \( \mathcal{R}^{II} \) through shadow observables could provide deeper insights into phase transitions and microscopic structures.  Therefore, we suggest further studying black hole shadow thermodynamics in rotating AdS spacetimes using alternative methods for projecting shadows at finite distances, such as those presented in \cite{ZhangGuo, Johannsen_2013, Cunha_2015, Bohn_2015, Cunha_2016}. These approaches could reveal whether black hole shadows encode thermodynamic properties, even in unusual scenarios, such as those beyond general relativity, in black mimickers, or in exotic compact objects \cite{Perlick_2022}.

\section{Conclusions}
 \label{sec:conclusions}
 
In this work, we investigated the thermodynamic properties, phase-transition structure, and microstructure of the Kerr-AdS black hole using the GTD framework and its shadow features. We treat the curvature radius $l$ as a thermodynamic variable, describing the black hole as a quasi-homogeneous system \cite{quevedo2019quasi}, consistent with our previous studies \cite{Ladino:2024ned,romero2024extended1}. In the canonical ensemble, for a slowly rotating black hole, we identify critical points in the equation of state at fixed $l$, consistent with prior findings \cite{wei2016analytical,gunasekaran2012extended,wei2021general}. Analyzing the Helmholtz free energy, we find that for $l > l_c$, the system undergoes a first-order small-to-large black hole phase transition, which becomes second-order at the critical value $l = l_c$.  This mirrors the liquid-gas transition in a van der Waals fluid but emerges independently of thermodynamic pressure. The phase structure also resembles that of RN-AdS black holes \cite{Ladino:2024ned}, suggesting an underlying universality class governing gravitational systems. In contrast, the grand-canonical ensemble exhibits a different phase structure, with an absence of the IBH phase and a single local minimum in the equation of state. Additionally, the system can transition into a globally preferred thermodynamic state, culminating in the well-known Hawking-Page phase transition, where the thermal rotating AdS phase \cite{caldarelli2000thermodynamics,hawking1999rotation} dominates over the LBH phase at sufficiently low temperatures. We propose a formula for the Hawking-Page temperature and, in the slow-rotation limit, derive a universal black hole ratio $\zeta = T_{HP}/T_{min}$, deviating from the well-known $\zeta=2/\sqrt{3}$ value for four-dimensional spherically symmetric black holes \cite{wei2020novel}. Notably, our results also differ from the previous work \cite{gao2022thermodynamics}, where treating the holographic central charge as a thermodynamic parameter leads to a homogeneous fundamental relation for the Kerr-AdS system.\\

Using the GTD framework and shadow analysis, we provide a new perspective on the phase structure and microstructure of the Kerr-AdS black hole, differing from Ruppeiner's approach. We analyze the thermodynamic equilibrium space of the Kerr-AdS black hole, capturing its phase structure in the canonical ensemble. The microstructural interactions of these rotating black hole resemble those of the RN-AdS black hole \cite{Ladino:2024ned}. In particular, superheated LBHs behave like an ideal gas with zero thermodynamic curvature, while SBHs predominantly exhibit repulsive interactions that can drive event horizon expansion, leading to a phase transition into a larger state. Given the absence of a fully consistent statistical description of black holes, we believe that our GTD-based characterization of phase transitions, expressed in terms of the curvature of the equilibrium space, provides an invariant approach for understanding the correspondence between black hole shadows, microstructure, and phase transitions. On the other hand, the shadow of Kerr-AdS black holes becomes asymmetric at higher \( a \) values, and a larger \( l \) causes the shadow to shrink, approaching the Kerr case. Applying the shadow thermodynamics formalism, we obtain asymmetric shadow profiles that reflect the thermodynamic properties of the black hole. We derive analytical expressions for the shadow and its critical quantities, including the critical celestial coordinates, which depend on the angular momentum \( J \) and observer distance \( r_0 \).   We also obtain constraints on the critical photon region \( r_{sc} \) and critical shadow radius \( R_{shc} \),  for \( 1.0692 \lesssim R_{shc} \lesssim 1.7043 \), the Kerr-AdS black hole can undergo a second-order phase transition. Our results confirm that the shadow radius \( R_{sh} \) and the shadow profiles effectively capture the phase structure of Kerr-AdS black holes, revealing both first-order and second-order phase transitions that align with corresponding thermodynamic response functions. Additionally, the phase structure in terms of \( R_{sh} \) is influenced by the curvature radius \( l \) but remains largely independent of the angular momentum \( J \), suggesting that spin is not a key thermodynamic factor. This supports the idea of the existence of a universality class, at least for electrovacuum black holes. Moreover, since the singularities of the curvature scalar in the GTD metric \( g^{II} \) in terms of the shadow coincide with the divergences in \( C_{lJ} \), the phase structure of the Kerr-AdS black hole can be inferred from shadow observations, suggesting that shadows encode both phase transitions and microstructural properties.\\

Future research may focus on studying the singularities at small entropy and curvature radius, using alternative shadow approaches to determine their physical meaning. Another direction is to explore the combined effects of charge \( Q \) and angular momentum \( J \) on the quasi-homogeneous GTD framework of Kerr-Newman AdS black holes \cite{caldarelli2000thermodynamics, Belhaj_2013, Peng}, in order to gain a deeper insight into general relativity. Additionally, investigating how extra dimensions affect phase structure and shadow thermodynamics, as well as extending the framework to phase transitions and microstructure in rotating black holes within alternative gravity theories, would further clarify the connection between shadows and thermodynamic properties in more complex scenarios \cite{belhaj2015ehrenfest, wei2020extended, Papnoi_2014, wei2016analytical, Xiong, Grenzebach1, Cunha_2016,  Kubizn_k_2007, Ishwaree, Nashed_2019, Capozziello_2019}. Finally, we can examine how current and future observational data from black hole shadows, such as those obtained by the next-generation Event Horizon Telescope \cite{EventHorizon1,EventHorizon2,EventHorizon3} or the Black Hole Explorer project \cite{Johnson}, might validate the predicted thermodynamic behaviors.

\let\oldaddcontentsline\addcontentsline
\renewcommand{\addcontentsline}[3]{}

\section*{Acknowledgments}
JML and CRF acknowledge support from Conahcyt-Mexico,  grants No. 4020764 and No. 4003366. This work was supported by UNAM-DGAPA-PAPIIT, grant No. 108225 and Conahcyt, grant No. CBF-2025-I-243.

\renewcommand{\addcontentsline}[3]{\oldaddcontentsline{#1}{#2}{#3}}
\appendix
\section{Explicit Formulas for Kerr-AdS Black Holes}\label{apenA}

The expressions for the entropy branches \( S_{\text{extr}} = S_{\text{extr}}(l, J) \), corresponding to the points where \( T_{\text{min,max}} = T(S_{\text{extr}}) \), as well as the critical values from Eq.~ \eqref{critical1}, defined by the conditions \( \partial_S T = \partial_{SS} T = 0 \), are consistently obtained up to order \( \mathcal{O}(J^2) \) as
\begin{equation}\label{sextrem}
\setlength{\jot}{0pt}
\begin{aligned}
  S_{extr}&=\frac{\pi}{6}\left(\sqrt{S_{1}}-l^2\pm\sqrt{9 l^4-S_{1}-\frac{8 l^2\left(108 J^2+l^4\right)}{\sqrt{S_{1}}}}\right),  \\
 S_{1}& \equiv S_{2}^{1 / 3} l^{4 / 3}+3 l^4+\frac{l^{8 / 3}\left(576 J^2+l^4\right)}{S_{2}^{1 / 3}},\\
 S_{2}& \equiv 93312 J^4+4320 J^2 l^4-l^8+72 \sqrt{2} J \sqrt{839808 J^6+59328 J^4 l^4+1686 J^2 l^8-l^{12}},
\end{aligned}
\end{equation}
and
\begin{equation}
\setlength{\jot}{0pt}
\begin{aligned}
 l_c &= \left(\frac{180701+29 \times 127 \sqrt{2689}}{3^3 \times 2^3}\right)^{1 / 4} \sqrt{J} \approx 6.4407 \sqrt{J}, \\
 S_c &=\sqrt{\frac{3 \times 61+3^2 \sqrt{2689}}{2^3}} \pi J\approx28.3114 J, \\
 M_c &=\left(\frac{106728859+7 \times 294029 \sqrt{2689}}{2^{17} \times 3^3 \times 5}\right)^{1 / 4} \sqrt{J} \approx 1.8637 \sqrt{J}, \\
 T_c &=\left(\frac{-100678099469+67 \times 47875921 \sqrt{2689}}{2^{29} \times 15^3 \times 11^2}\right)^{1 / 4} \frac{1}{\pi \sqrt{J}}\approx\frac{0.0419}{\sqrt{J}}, \\
 r_{hc}&=\left(\frac{-3 \times 172171+21 \times 541 \sqrt{2689}}{11^2 \times 2^3}\right)^{1 / 4} \sqrt{J}\approx2.9430 \sqrt{J}.
\end{aligned}
\label{critical1apen}
\end{equation}
The explicit forms of the second derivatives of the thermodynamic potential of Eq.~ \eqref{fundame1} are
\begin{align}
    M_{,SS} &= 
    \frac{
        - S^4(l^2 \pi - 3 S) (l^2 \pi + S)^3 
        + 24 J^2 l^2 \pi^3 S^2 (l^2 \pi + S)^2 (l^2 \pi + 2 S)
        + 16 J^4 l^6 \pi^7 (3 l^2 \pi + 4 S)
    }{
        64 l^8 \pi^{6} S^4 
         M^3
    },\nonumber \\
    M_{,JJ} &= 
    \frac{
         (l^2 \pi + S)^3
    }{
        4l^6 \pi^{3} M^3
    },\nonumber \\
    M_{,ll} &= 
    \frac{
        18 J^2 l^2 \pi^3 S^2 (l^2 \pi + S)^2 + 3 S^4 (l^2 \pi + S)^3 + 
        8 J^4 l^4 \pi^6 (3 l^2 \pi + 2 S)
    }{
        8Sl^{10} \pi^{6} M^3
    },\nonumber \\
    M_{,SJ} &= 
    -\frac{J  (l^2 \pi + S) \big[ 4 J^2 l^4 \pi^4 + 3 S^2 (l^2 \pi + S)^2 \big]}{
        8l^6 S^3  \pi^{3} M^3
    }, \label{second_deriv}\\
    M_{,Sl} &= -
    \frac{
        8 J^4 l^6 \pi^7 +18 J^2 l^2 \pi^3 S^2 (l^2 \pi + S)^2 + 
        3 S^4 (l^2 \pi + S)^3
    }{
        16 l^9 \pi^{6} S^2 M^3
    }, \nonumber\\
    M_{,Jl} &= 
    -\frac{ J^3 (l^2 \pi + S)}{
        S l^5 M^3
    }.\nonumber
\end{align}
Furthermore, using these second derivatives along with the Nambu bracket notation \cite{mansoori2015hessian}, we can compute the relevant response functions within a thermodynamic equilibrium space characterized by the coordinates \( (S, l, J) \), as

\setlength{\jot}{10pt}  

\begin{align}
\kappa_{S\Omega} &\equiv \left(\frac{\partial l}{\partial \Psi}\right)_{S\Omega} = \frac{\{S,l,\Omega\}}{\{S,\Psi,\Omega\}} = \frac{M_{,JJ}}{M_{,ll}M_{,JJ} - (M_{,Jl})^2}, \label{kappasomega}\\
\kappa_{Sl} &\equiv \left(\frac{\partial J}{\partial \Omega}\right)_{Sl} = \frac{\{S,l,J\}}{\{S,l,\Omega\}} = \frac{1}{M_{,JJ}}, \\
\kappa_{SJ} &\equiv \left(\frac{\partial l}{\partial \Psi}\right)_{SJ} = \frac{\{S,J,l\}}{\{S,J,\Psi\}} = \frac{1}{M_{,ll}}, \\
\alpha_{Sl} &\equiv \left(\frac{\partial J}{\partial T}\right)_{Sl} = \frac{\{S,l,J\}}{\{T,S,l\}} = \frac{1}{M_{,SJ}}, \\
\alpha_{SJ} &\equiv \left(\frac{\partial l}{\partial T}\right)_{SJ} = \frac{\{S,l,J\}}{\{T,S,J\}} = \frac{1}{M_{,Sl}}, \\
C_{lJ} &\equiv T \left(\frac{\partial S}{\partial T}\right)_{lJ} = T \frac{\{S,l,J\}}{\{T,l,J\}} = \frac{T}{M_{,SS}}, \\
C_{\Psi \Omega} &\equiv T \left(\frac{\partial S}{\partial T}\right)_{\Psi \Omega} = T \frac{\{S,\Psi,\Omega\}}{\{T,\Psi,\Omega\}} \notag \\
&= \frac{T \big[M_{,ll}M_{,JJ} - (M_{,Jl})^2\big]}{M_{,SS} \big[M_{,ll}M_{,JJ} - (M_{,Jl})^2\big] - (M_{,Sl})^2 M_{,JJ} - (M_{,SJ})^2 M_{,ll} + 2M_{,Sl}M_{,Jl}M_{,SJ}}, \\
C_{\Omega l} &\equiv T \left(\frac{\partial S}{\partial T}\right)_{\Omega l} = T \frac{\{S,l,\Omega\}}{\{T,l,\Omega\}} = \frac{T M_{,JJ}}{M_{,SS}M_{,JJ} - (M_{,SJ})^2}; \label{clw}
\end{align}
where \( C_{xy} \), \( \kappa_{xy} \), and \( \alpha_{xy} \) represent the heat capacity, the compressibility parameter, and the coefficient of thermal expansion, respectively, at a fixed set of thermodynamic parameters $(x,y)$ \cite{Ladino:2024ned}.\\

Taking \( S_l \equiv l^2 \pi + S \), the explicit form of the Ricci scalar \( \mathcal{R}^{II} \) for the metric \( g^{II} \) in Eq.~ \eqref{metric2} is
\begin{align}
\mathcal{R}^{II}=\mathcal{R}^{II}(S,l,J)=\frac{\mathcal{N}^{II}}{\mathcal{D}^{II}},
\end{align}
where
\begin{equation}
\setlength{\jot}{0pt}
\begin{aligned}
& \mathcal{N}^{II}=-8 l^4 \pi^3\left(-4096 J^{14} l^{18} \pi^{23}\left(3 l^2 \pi+4 S\right)\left(9 l^4 \pi^2+16 l^2 \pi S+8 S^2\right)-\right. \notag\\
& 9 J^2 l^2 \pi^3 S^{11}S_l^9\left(207 l^6 \pi^3+322 l^4 \pi^2 S+105 l^2 \pi S^2+30 S^3\right)+ \notag\\
& 3 S^{13}S_l^{10}\left(-14 l^6 \pi^3+10 l^4 \pi^2 S+111 l^2 \pi S^2+99 S^3\right)+ \notag\\
& 1024 J^{12} l^{14} \pi^{19} SS_l^2\left(162 l^8 \pi^4+594 l^6 \pi^3 S+721 l^4 \pi^2 S^2+260 l^2 \pi S^3-48 S^4\right)- \notag\\
& 24 J^4 I^4 \pi^6 S^9S_l^7\left(2011^8 \pi^4+511^6 \pi^3 S-3681^4 \pi^2 S^2-4 l^2 \pi S^3+204 S^4\right)+ \notag\\
& 32 J^6 l^6 \pi^9 S^7S_l^6\left(1155 l^8 \pi^4+5691 l^6 \pi^3 S+8482 l^4 \pi^2 S^2+5010 l^2 \pi S^3+1125 S^4\right)+ \notag\\
& 256 J^{10} 1^{10} \pi^{15} S^3S_l^3\left(1305 l^{10} \pi^5+5913 l^8 \pi^4 S+10569 l^6 \pi^3 S^2+9160 l^4 \pi^2 S^3+3612 l^2 \pi S^4+432 S^5\right)+ \notag\\
&\left.64 J^8 l^8 \pi^{12} S^5S_l^4\left(3204 l^{10} \pi^5+15732 l^8 \pi^4 S+30599 l^6 \pi^3 S^2+30134 l^4 \pi^2 S^3+15177 l^2 \pi S^4+3096 S^5\right)\right)\notag,\\\\
&\mathcal{D}^{II}=\mathrm{\beta_M}S_l\left(-3\left(I^2 \pi-3 S\right) S^7S_l^6-128 J^8 l^{10} \pi^{13}\left(3 l^2 \pi+4 S\right)+6 J^2 l^2 \pi^3 S^5S_l^5\left(11 l^2 \pi+27 S\right)+\right. \notag\\
&\left.96 J^6 l^6 \pi^9 SS_l^2\left(3 I^4 \pi^2+2 l^2 \pi S-4 S^2\right)+8 J^4 l^4 \pi^6 S^3S_l^3\left(36 l^4 \pi^2+79 l^2 \pi S+33 S^2\right)\right)^2 \notag .
\end{aligned}
\end{equation}
Finally, by considering the critical values \( r_{sc} \equiv \bar{r} \sqrt{J} \), \( l_c  \equiv \bar{l} \sqrt{J} \), and \( M_c \equiv \bar{M} \sqrt{J} \), and substituting the critical impact parameters from Eqs.~ \eqref{criticalimpact1} and \eqref{criticalimpact2} into Eqs.~ \eqref{celestial1} and \eqref{celestial2}, we obtain (for $\theta_0=\pi/2$) the explicit expressions for the critical celestial coordinates of the shadow as
\begin{align}
& \sin \rho_c  =  -\frac{\mathcal{C}}{\left(1-\bar{l}^2 \bar{M}^2\right) \sqrt{\mathcal{A}}} \sqrt{\frac{\left(1-2 \bar{l}^2 \bar{M}^2\right)(\mathcal{A}+\mathcal{B})}{\mathcal{C}^2}}  , \label{critical2}\\
& \sin \sigma_c  = \frac{1-\bar{l}^2 \bar{M}^2}{\bar{l}^4 \bar{M}^{5 / 2} J} \sqrt{\frac{\mathcal{D}}{-\mathcal{A}-\mathcal{B}}}\left(\sqrt{\frac{\mathcal{A} \bar{l}^4 \bar{M}^4}{2 \bar{l}^2 \bar{M}^2-1}}-\mathcal{C} r_0^2\right)  , \label{critical3}
\end{align}
\begin{equation}
\setlength{\jot}{0pt}
\begin{aligned}
 \text{with} \quad\quad \mathcal{A} \equiv & -\frac{1-2 \bar{l}^2 \bar{M}^2}{\bar{l}^4 \bar{M}^4}\Big\{\bar{r}\left[-3 \bar{r}+\bar{l}^2 \bar{M} \bar{r}(6 \bar{M}+\bar{r})+\bar{l}^4 \bar{M}\left(2+\bar{M}^2 \bar{r}(\bar{r}-3 \bar{M})\right)\right] J+\mathcal{C} r_0^2\Big\}^2 , \notag \\
 \mathcal{B} \equiv & -4 \bar{r}^2\left[\bar{l}^4 \bar{M}-\left(1-\bar{l}^2 \bar{M}^2\right)^2 \bar{r}-\bar{l}^2 \bar{M}^3 \bar{r}^4+\bar{r} \mathcal{C}\right] \mathcal{D} , \notag \\
 \mathcal{C} \equiv & -1+\bar{l}^2 \bar{M}\Big\{\bar{r}+\bar{M}\left[2+2 \bar{M} \bar{r}^3+\bar{l}^2 \bar{M}(\bar{r}-\bar{M})\right]\Big\}, \label{Cequation} \\
 \mathcal{D} \equiv &  \bar{l}^4 \bar{M} J^2-2\left(-1+\bar{l}^2 \bar{M}^2\right)^2 J^{3 / 2} r_0+\bar{l}^2 \bar{M}\left(1+\bar{l}^2 \bar{M}^2\right) J r_0^2+\bar{l}^2 \bar{M}^3 r_0^4 .  \notag
\end{aligned}
\end{equation}
Thus, the critical celestial coordinates of the shadow, \(\rho_c\) and \(\sigma_c\), depend only on \(\{J, r_0\}\).

%
 

\let\oldaddcontentsline\addcontentsline
\renewcommand{\addcontentsline}[3]{}
 \setlength{\bibsep}{0pt}
\bibliographystyle{unsrt}


\end{document}